\documentclass[12pt,a4paper]{article}

\usepackage{amsmath,amssymb,amsthm,amsfonts}
\usepackage{bm}
\usepackage{graphicx}
\usepackage{hyperref}
\usepackage{geometry}
\usepackage{color}
\usepackage{subcaption}
\usepackage{cite}
\usepackage{ulem}

\numberwithin{equation}{section}

\begin{document}
\begin{center}
{\large \bf
Disentangling new physics with quantum entanglement \\[5mm]
in $t\bar{t}$ production at future lepton colliders 
}
\vskip 1.0cm
{\large
Masato Arai$^a$\footnote{arai(at)sci.kj.yamagata-u.ac.jp}, 
Kentarou Mawatari$^{b,c,d}$\footnote{mawatari(at)iwate-u.ac.jp},
Nobuchika Okada$^e$\footnote{okadan(at)ua.edu}, 

}
\vskip 1.0cm
  {\it
  $^a$Faculty of Science, Yamagata University, Yamagata 990-8560, Japan \\
  \vskip 0.2cm
  $^b$Faculty of Education, Iwate University, Morioka, Iwate 020-8550, Japan \\
  $^c$Graduate School of Arts and Sciences, Iwate University,\\ Morioka, Iwate 020-8550, Japan \\
  $^d$Graduate School of Science and Engineering, Iwate University,\\ Morioka, Iwate 020-8550, Japan\\
  \vskip 0.2cm
  $^e$Department of Physics and Astronomy \\
  University of Alabama, Tuscaloosa, AL35487, USA
}
\vskip 1.5cm
\end{center}

\begin{abstract}
We investigate quantum entanglement and Bell-inequality violation in top-antitop pair production at future lepton 
 colliders such as the International Linear Collider (ILC) and multi-TeV muon colliders.
Within the Standard Model (SM), the process proceeds through $s$-channel $\gamma$ and $Z$ exchange and 
 exhibits characteristic spin-correlation patterns that encode a non-trivial amount of entanglement.
We then examine how these features are modified in several well-motivated extensions of the SM:
 (i) a neutral scalar mediator that couples to charged leptons and top quarks via Yukawa interactions and contributes as 
 an additional $s$-channel exchange; (ii) the minimal gauged $U(1)_{B-L}$ model, which introduces a new neutral 
 gauge boson $Z'$ coupling vectorially to SM fermions; and (iii) a Randall-Sundrum scenario, in which the exchange 
 of massive Kaluza-Klein gravitons arising from a warped extra dimension induces additional spin-dependent interactions.
For all cases, we evaluate quantum-information observables including the entanglement marker, the concurrence, 
 and the maximal Clauser-Horne-Shimony-Holt parameter, and study their dependence on the center-of-mass 
  energy, scattering angle, and model parameters.
We find that, relative to the SM expectation, the entanglement is typically reduced in the scalar-mediator scenario, 
 while sizable deviations can arise in the $U(1)_{B-L}$ and Randall-Sundrum cases for phenomenologically relevant 
 regions of parameter space.
These results demonstrate the potential of quantum-information observables as sensitive probes of new 
 particles and their interaction patterns
 in future lepton colliders.
 \end{abstract}

\newpage

\section{Introduction}
\label{sec:intro}

Quantum entanglement is one of the most fundamental features of quantum mechanics.
It refers to the phenomenon in which multiple quantum systems form a single global quantum state that cannot be 
 described as a product of independent states of its subsystems.
Although physical observables are measured locally on each subsystem, the measurement outcomes exhibit strong 
 correlations that cannot be explained within classical physics.
Such non-classical correlations originate from the global structure of the quantum state itself.
Well-known low-energy realizations of quantum entanglement include entangled photon pairs~\cite{Bell:1964kc,Aspect:1981nv} 
 and ultracold atomic systems~\cite{Bloch:2012uep}, where highly controlled experimental environments have enabled precise 
  tests of entanglement generation and its properties.

Quantum entanglement, however, is a universal phenomenon that arises in all systems governed by quantum mechanics and 
 is by no means restricted to low-energy physics.
In particular, heavy particles produced in high-energy processes provide a unique arena for studying entanglement: due to the 
 short-distance nature of the interactions and the rapid decay of unstable particles, phase information and spin correlations of the 
  quantum state can be preserved and accessed experimentally.
From this perspective, high-energy scattering processes offer a new experimental and theoretical stage for exploring quantum 
 entanglement.

Indeed, correlations among spin degrees of freedom in particle-antiparticle pair production have long been studied in 
 high-energy physics.
In processes such as top-quark pair production and electroweak reactions, the spin states of the produced particles can be 
 reconstructed from their decay products, enabling systematic analyses of spin correlations within the density-matrix formalism~\cite{Kane:1991bg,Bernreuther:1993hq,Parke:1996pr,Bernreuther:1997gs,Bernreuther:2004jv,Uwer:2004vp,Baumgart:2012ay,Bernreuther:2015yna}.
These spin correlations reflect the underlying interaction dynamics of the system and describe quantum correlations in a broad sense, 
 not necessarily limited to quantum entanglement.
Experimentally, spin correlations in top-antitop production have already been measured at the Tevatron in proton-antiproton 
 collisions by the D0 and CDF collaborations~\cite{CDF:2010yag,D0:2011rkb,D0:2015kta}, and subsequently at the Large Hadron 
  Collider (LHC) in proton-proton collisions by the ATLAS and CMS 
  experiments~\cite{ATLAS:2012ao,CMS:2013roq,ATLAS:2014abv,CMS:2019nrx,ATLAS:2019zrq}.

Building on this progress, quantum correlations in top-antitop production at the
 LHC, including quantum entanglement and Bell-inequality violation, have attracted
 considerable attention in recent years.
Using Run-2 LHC data, Afik et al.~\cite{Afik:2020onf} demonstrated that quantum
 entanglement in $t\bar t$ pairs can be observed with high statistical significance,
 marking the first proposal to measure entanglement in a quark-antiquark system
 and the highest-energy realization of such a study to date.
Subsequent works have systematically investigated both the structure of
 entanglement and the conditions for Bell-inequality violation in $t\bar t$
 production at the LHC, as well as their experimental accessibility~\cite{Fabbrichesi:2021npl, Severi:2021cnj, Afik:2022kwm, Aguilar-Saavedra:2022uye, Maltoni:2024tul}.
Related analyses have also been extended to other massive systems at colliders,
such as spin-1 $W^+W^-$ pairs~\cite{Barr:2021zcp, Fabbrichesi:2023cev}.
Importantly, recent measurements by ATLAS and CMS~\cite{ATLAS:2023fsd,CMS:2024zkc} have 
 provided the first experimental evidence for entanglement in $pp \to t\bar t$, demonstrating the 
 observability of quantum correlations in the SM at high energies.

While top-antitop production has been extensively studied at hadron colliders
 such as the LHC, where a wide variety of quantum-mechanical observables have been
 analyzed, complementary studies have also been carried out at lepton colliders.
Lepton colliders provide an exceptionally clean experimental environment, with
 precisely controlled initial states and production mechanisms dominated by
 electroweak interactions rather than strong dynamics.
These features make them an ideal setting for high-precision investigations of
 spin correlations and quantum correlations in heavy-particle production.

Motivated by these advantages, detailed analyses of quantum spin observables in
 top-quark pair production at lepton colliders have been performed in recent
 years, clearly demonstrating their potential and relevance~\cite{Maltoni:2024csn}.
The clean environment and tunable kinematics of lepton colliders allow for a
 transparent interpretation of spin-correlation effects and provide a natural
 laboratory for studying quantum entanglement in $t\bar t$ systems.

Similar quantum-information-oriented analyses have also been
 carried out for other fermion-antifermion production processes at lepton
 colliders.
In particular, systematic studies of spin correlations, quantum entanglement,
and Bell-inequality violation in $\tau^+\tau^-$ production at
 electron-positron colliders have been proposed~\cite{Ehataht:2023zzt}.
Recent works have further investigated the sensitivity of such measurements at facilities 
such as Beijing Electron-Positron Collider \cite{Han:2025ewp} and the proposed
 Super Tau-Charm Facility \cite{Yang:2026uwu}.
Tau-lepton pairs provide a well-controlled system for testing
 quantum correlations and for comparing how different fermion species and
 interaction structures influence their manifestation.
More generally, quantum correlations in fermion-pair production have been
analyzed using density-matrix and quantum-tomography formulations,
including processes such as $f\bar f \to l^+l^-$ in the SM and its
SMEFT extensions, with particular attention to the phenomenologically
relevant case $l=\tau$~\cite{LoChiatto:2024dmx}.
 
More general and systematic approaches to quantum correlations in
high-energy processes have recently been developed.
Early progress in this direction explored quantum tomography of collider
states, including approaches based on decay observables as well as
production kinematics alone~\cite{Cheng:2024rxi}.
More recently, a fully automated framework was presented to construct
 production spin-density matrices for generic collider processes~\cite{Durupt:2025wuk}.
A particularly comprehensive theoretical framework for quantum
correlations in fermion-pair production at lepton colliders,
including spin correlations, entanglement, and Bell nonlocality,
was developed in Ref.~\cite{Altakach:2026fpl}, where analytic expressions
for the corresponding spin-density matrices
were derived in a general setup with polarized beams and effective
interactions.
These theoretical approaches are complementary to quantum-tomography methods,
which aim at reconstructing the same density matrix from experimental data.
The choice of spin basis is also important for extracting quantum
correlations, and optimized bases have been investigated for
enhancing entanglement and Bell-inequality violation in top-antitop
events at hadron and future lepton colliders~\cite{Cheng:2024btk}.
Building on these developments, a number of works have performed detailed analyses of
quantum-information observables such as concurrence and Bell parameters
in processes involving fermion pairs~\cite{Zhang:2026nwm, Guo:2026yhz},
 while other measures of quantum correlations beyond entanglement, such as
 quantum discord, have also been investigated in collider processes~\cite{Han:2024ugl}.
In a different but related direction, quantum-information aspects of
top-quark systems have also been explored in bound-state configurations,
such as toponium~\cite{Antozzi:2026vdi}.
The potential of quantum-entanglement observables for discriminating
different classes of new physics in fermion-pair production has also
been investigated \cite{Duch:2024pwm}.

In this work, we perform a comprehensive analysis of quantum entanglement in
 $t\bar t$ production at lepton colliders such as the International Linear Collider (ILC) and multi-TeV muon colliders, 
  within the Standard Model (SM) and three
 representative scenarios involving neutral $s$-channel mediators:
(i) a neutral scalar mediator $\Phi$ that couples to charged leptons
and top quarks
through Yukawa interactions;
(ii) the minimal gauged $U(1)_{B-L}$ model
 ~\cite{Mohapatra:1980qe, Marshak:1979fm, Wetterich:1981bx, Masiero:1982fi, Mohapatra:1982xz, Buchmuller:1991ce}, featuring a new vector boson $Z'$
that couples vectorially to SM fermions; and
(iii) the Randall-Sundrum (RS) model~\cite{Randall:1999ee}, in which Kaluza-Klein (KK) excitations of the
graviton contribute as spin-2 $s$-channel mediators.
These scenarios provide representative examples of scalar, vector, and tensor interactions.
Our goal is not merely to quantify the amount of entanglement generated in each model, but to investigate 
whether quantum-information observables can reveal characteristic signatures of 
 new paticles and their interaction patterns.
By treating the SM and all new-physics scenarios on equal footing, 
 we assess differences in entanglement and Bell-nonlocality observables among the SM and the three new-physics 
 scenarios.
We show that these observables provide additional information 
complementary to conventional classical observables such as total and differential cross sections.

The remainder of this paper is organized as follows.
In Sec.~\ref{sec:formalism} we present the general density-matrix framework for
 two spin-$\tfrac{1}{2}$ particles and define the quantum-information observables
 used in our analysis, including entanglement markers, concurrence and Bell-type quantities.
The new physics scenarios considered in this work are summarized in
 Sec.~\ref{sec:models}.
In Sec.~\ref{sec:amplitudes} we derive the helicity amplitudes for
 $l^-l^+\to t\bar t$ $(l=e,\mu)$ in the SM and in the benchmark extensions.
Numerical results for entanglement marker, concurrence and Bell-inequality violation at future lepton
 colliders are presented in Sec.~\ref{sec:numerics}, and conclusions are given in
 Sec.~\ref{sec:conclusion}. 
Appendix \ref{app:chiral_vector} presents an analytic study of a generic chiral gauge interaction, clarifying 
 the dependence of the entanglement observables on the chiral couplings.
Appendix \ref{app:vector_interference} investigates the interference between chiral and vector-like interactions, providing 
 an analytic interpretation
 of the energy-dependent evolution of the entanglement extrema and
 the weakened-entanglement regions observed in the $U(1)_{B-L}$ scenario.
Appendix \ref{app:RS_angular} demonstrates that the disconnected entangled regions in the RS model originate from the 
 characteristic angular dependence due to a single spin-2 KK graviton exchange.

\section{Formalism: Quantum-information observables}
\label{sec:formalism}

\subsection{Spin density matrix and correlation coefficients}
\begin{figure}[htbp]
    \centering
    \includegraphics[width=0.3\textwidth]{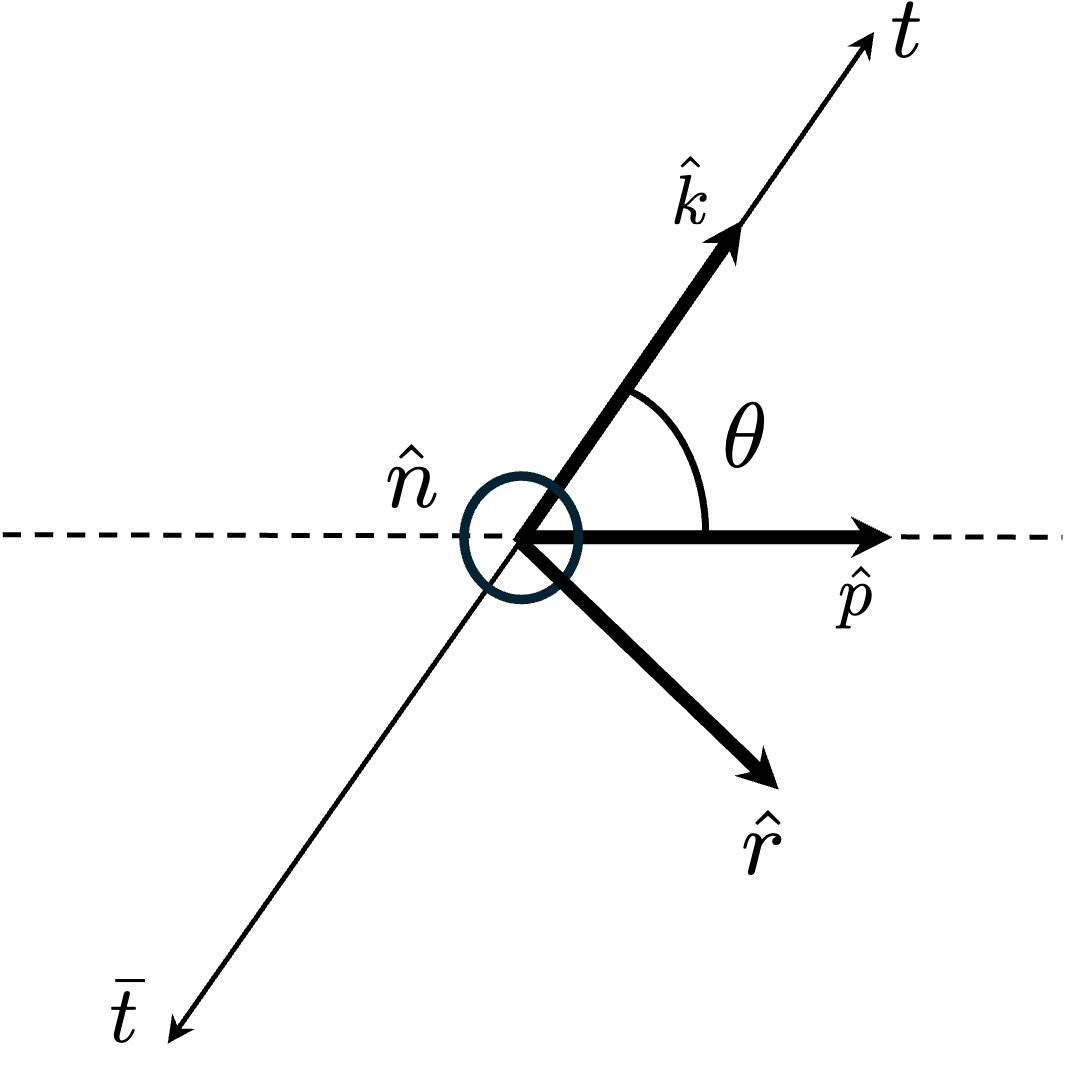}
\caption{Orthonormal basis $(\hat r,\hat n,\hat k)$ in the $t\bar t$ rest frame, 
defined with respect to the top-quark direction $\hat k$ and beam direction $\hat p$.}
\label{fig:frame}
\end{figure}

Spin correlations in a top-antitop system can only be fully specified once nine independent parameters 
 are fixed, as is the case for any two-qubit state. 
A convenient way to encode these degrees of freedom is through the standard decomposition of the spin density matrix,
\begin{equation}
\rho=\frac14\left(
\mathbf 1 \otimes \mathbf 1
+\mathcal B_{1}\cdot\boldsymbol{\sigma}\otimes\mathbf 1
+\mathcal B_{2}\cdot\mathbf 1\otimes\boldsymbol{\sigma}
+\mathcal C\cdot\boldsymbol{\sigma} \otimes \boldsymbol{\sigma}
\right) ,\label{rho}
\end{equation}
where the vectors $\mathcal B_1 = (B_{1i})$ and 
 $\mathcal B_2 = (B_{2i})$, with $i=1,2,3$, quantify, respectively, 
 the top and antitop polarizations along the spatial axes, 
 and $\mathcal C = (C_{ij})$, with $i,j=1,2,3$, 
 contains nine parameters and encodes all possible spin-spin correlations. 
Here, $\boldsymbol{\sigma} = (\sigma^1,\sigma^2,\sigma^3)$ denotes the Pauli matrices. 
To give operational meaning to these components, one must choose a spatial triad.
In what follows we adopt the customary 
 orthonormal basis built from the top-quark momentum direction,
\begin{equation}
\hat k=\text{top direction},\qquad
\hat r=\frac{\hat p-\hat k\cos\theta}{\sin\theta},\qquad
\hat n=\frac{\hat p\times\hat k}{\sin\theta},
\label{helbasis}
\end{equation}
with $\hat p$ the incoming $l^-$
direction, and with $\theta$ defined through $\cos\theta=\hat k \cdot \hat p$ in the $t\bar t$ 
rest frame.
The geometric configuration of these vectors is depicted in Fig.~\ref{fig:frame}.


\subsection{Entanglement measures and Bell inequality}
To quantify bipartite entanglement between the two spin-$1/2$ systems, 
 we use the concurrence $C$ introduced by Wootters \cite{Wootters:1997id}.
For a given two-qubit density matrix $\rho$, we define the
 spin-flipped density matrix as
\begin{equation}
  \tilde{\rho} = (\sigma_y \otimes \sigma_y)\, \rho^{*}\, (\sigma_y \otimes \sigma_y),
\end{equation}
where the asterisk denotes complex conjugation in the computational basis.
The non-Hermitian matrix
\begin{equation}
  R = \sqrt{\sqrt{\rho}\tilde{\rho} \sqrt{\rho}}
\end{equation}
has four real, non-negative eigenvalues $\lambda_i$ ($i=1,\dots,4$), which we
order such that $\lambda_1\ge \lambda_2\ge \lambda_3\ge \lambda_4$.
The concurrence is then given by
\begin{equation}
  C = \max\left\{0,
    \lambda_1 - \lambda_2
    - \lambda_3 - \lambda_4
  \right\}.
  \label{eq:concurrence-general}
\end{equation}
It satisfies $0\le C\le 1$, with $C=0$ if and only if the state is separable,
and $C=1$ for maximally entangled Bell states.

In collider environments, however, $C$ is not an ideal observable. 
Extracting them from data typically inflates the impact of statistical fluctuations; related estimators have 
 been shown to suffer from an intrinsic upward bias~\cite{Severi:2021cnj}. 
For this reason, it is advantageous to work with alternative quantities that retain sensitivity to quantum 
 correlations while being more robust experimentally. 
As established in Refs.~\cite{Afik:2022kwm, Aguilar-Saavedra:2022uye, Maltoni:2024tul}, 
the so-called entanglement markers 
can be built from the diagonal entries of~$\mathcal C$:
\begin{align}
D^{(1)} &= \frac{1}{3} (C_{kk}+C_{rr}+C_{nn}), \label{D1}\\
D^{(k)} &= \frac{1}{3} (C_{kk}-C_{rr}-C_{nn}), \\
D^{(r)} &= \frac{1}{3} (-C_{kk}+C_{rr}-C_{nn}), \\
D^{(n)} &= \frac{1}{3} (-C_{kk}-C_{rr}+C_{nn}). \label{D4}
\end{align}
Here, the indices $i,j=1,2,3$ correspond to the orthonormal basis
$(\hat r,\hat n,\hat k)$, respectively.
A convenient sufficient criterion for entanglement in \eqref{rho} is
\begin{equation}
D_{\min}\equiv \min\{D^{(1)},D^{(k)},D^{(r)},D^{(n)}\} < -\frac{1}{3}.
\label{Dentanglement}
\end{equation}
This inequality underlies the recent experimental demonstrations of top-quark entanglement in $pp$ 
 collisions~\cite{ATLAS:2023fsd, CMS:2024zkc}, where $D^{(1)}$ becomes the smallest of the four markers near threshold.

While \eqref{Dentanglement} is normally only sufficient, it becomes necessary when both top and antitop polarizations 
 vanish, $\mathcal B_1=\mathcal B_2=0$, and $\mathcal C$ is diagonal. In that case, the concurrence reduces to
\begin{equation}
C=\tfrac12 \max\bigl(0,-1-3D_{\min}\bigr),
\end{equation}
which makes it explicit that nonzero entanglement appears precisely when $D_{\min}<-\tfrac13$.

The nonlocal properties of the two-qubit state $\rho$ can be probed by Bell inequalities.
The simplest and most widely studied version is the Clauser-Horne-Shimony-Holt (CHSH) inequality~\cite{Clauser:1969ny}, 
 formulated in terms of spin measurements performed independently on the two qubits.
One may imagine two distant observers, Alice and Bob, who each choose between two possible spin-measurement 
 directions.
Let Alice's choice of axes be the unit vectors $\hat a$ and $\hat a'$, and Bob's be $\hat b$ and $\hat b'$.
For a given pair of measurement directions, the spin-spin correlation is defined as
\begin{equation}
E(\hat a,\hat b)
= \mathrm{Tr}\left[\rho (\hat a\cdot\boldsymbol{\sigma}) \otimes (\hat b\cdot\boldsymbol{\sigma})\right],
\end{equation}
where $\boldsymbol{\sigma}=(\sigma_x,\sigma_y,\sigma_z)$ are the Pauli matrices.
The CHSH Bell parameter $B$ is constructed as
\begin{equation}
B = E(\hat a,\hat b) + E(\hat a,\hat b')
+ E(\hat a',\hat b) - E(\hat a',\hat b') .
\label{eq:CHSH-definition}
\end{equation}
Local hidden-variable theories impose the CHSH bound~\cite{Bell:1964fg}
\begin{equation}
|B| \le 2 ,
\end{equation}
whereas quantum mechanics allows values up to the Tsirelson limit $2\sqrt{2}$~\cite{Cirelson:1980ry}.
This inequality thus provides a sharp criterion for diagnosing nonclassical correlations in a two-qubit system.

For an arbitrary mixed two-qubit state, the maximal quantum violation $B_{\max}$ obtainable by optimizing over all measurement directions can be expressed entirely in terms of the correlation matrix $\mathcal{C}$ defined in Eq.~\eqref{rho}.
Introducing
\begin{equation}
U = \mathcal{C}^{\mathrm T} \mathcal{C},
\end{equation}
and denoting its eigenvalues by $u_1 \ge u_2 \ge u_3 \ge 0$, the Horodecki criterion~\cite{Horodecki:1995nsk} states
\begin{equation}
B_{\max} = 2\sqrt{u_1 + u_2},
\label{eq:Bmax}
\end{equation}
and CHSH violation occurs if and only if
\begin{equation}
B_{\max} > 2
\qquad\Longleftrightarrow\qquad
u_1 + u_2 > 1.
\end{equation}

In collider applications, $B_{\max}$ provides a clean and basis-independent diagnostic of the quantum nonlocality present in the spin state of the produced pair.
The correlation matrix $\mathcal{C}$ can be reconstructed experimentally from angular distributions in the decay products of the top and antitop quarks, 
 as discussed for example in~\cite{Severi:2021cnj, Afik:2022kwm,Aguilar-Saavedra:2022uye}.
In the present theoretical work, we compute $\mathcal{C}$ and consequently $B_{\max}$ directly from the density matrix~\eqref{rho}.

\subsection{Construction of the spin density matrix from helicity amplitudes}
\label{subsec:rho-from-amps}

We now specialize the general two-qubit formalism to the process
 $l^-l^+\to t\bar t$ $(l=e,\mu)$ in the helicity basis.
Throughout this paper, helicities are denoted by $\pm$, corresponding to
$\pm\tfrac12$.

We denote the helicity amplitudes for this process by
 $\mathcal{M}(\lambda_{l^-}, \lambda_{l^+}; \lambda_t, \lambda_{\bar t})$,
 where $\lambda_{l^-}$, $\lambda_{l^+}$, $\lambda_t$, and $\lambda_{\bar t}$
 represent the helicities of the incoming lepton, incoming antilepton,
 top quark, and antitop quark,
 respectively.

Since we consider unpolarized initial leptons, we start directly from the
spin-averaged quantity.
We define the (unnormalized) spin matrix of the final $t\bar t$ system as
\begin{equation}
  R_{\lambda_t\lambda_{\bar t},\,\lambda_t'\lambda_{\bar t}'}
  \equiv
  \frac{1}{4}
  \sum_{\lambda_{l^-},\lambda_{l^+}}
  \mathcal{M}(\lambda_{l^-},\lambda_{l^+};\lambda_t,\lambda_{\bar t})\,
  \mathcal{M}^{*}(\lambda_{l^-},\lambda_{l^+};\lambda_t',\lambda_{\bar t}') ,
  \label{eq:R-from-amps-sec2}
\end{equation}
where $(\lambda_t,\lambda_{\bar t})$ and $(\lambda_t',\lambda_{\bar t}')$ label
the row and column of the $4\times4$ matrix $R$.

The normalized spin density matrix relevant for entanglement observables is
obtained by dividing by the spin-summed rate,
\begin{equation}
  \rho_{\lambda_t\lambda_{\bar t},\,\lambda_t'\lambda_{\bar t}'}
  =
  \frac{
    R_{\lambda_t\lambda_{\bar t},\,\lambda_t'\lambda_{\bar t}'}
  }{
    \displaystyle
    \sum_{\lambda_t,\lambda_{\bar t}}
    R_{\lambda_t\lambda_{\bar t},\,\lambda_t \lambda_{\bar t}}
  } \,,
  \label{eq:rho-from-R-sec2}
\end{equation}
so that $\mathrm{Tr}\,\rho=1$ by construction.

Once $\rho$ is obtained, we decompose it in the Pauli basis as in Eq.~\eqref{rho}.
The coefficients $B_{1i}, B_{2i}$ and $C_{ij}$ are extracted from $\rho$ via
\begin{align}
  B_{1i} &= \mathrm{Tr}\!\left[\rho\,(\sigma_i\otimes \mathbf 1)\right], &
  B_{2j} &= \mathrm{Tr}\!\left[\rho\,(\mathbf 1\otimes \sigma_j)\right], &
  C_{ij} &= \mathrm{Tr}\!\left[\rho\,(\sigma_i\otimes \sigma_j)\right].
  \label{eq:BC-from-rho-sec2}
\end{align}

For practical calculations, it is convenient to write the density matrix
explicitly in the helicity basis of the $t\bar t$ system.
We choose the spin quantization axis along the top-quark momentum.
We order the basis states as
\[
|1\rangle = |+,-\rangle ,\qquad
|2\rangle = |+,+\rangle ,\qquad
|3\rangle = |-,-\rangle ,\qquad
|4\rangle = |-,+\rangle ,
\]
where the first (second) entry denotes the helicity of the top (anti-top)
 quark. 
Since the spin quantization axis is defined by the top-quark momentum,
 the helicity of the anti-top quark is evaluated with respect to the same axis.
As a result, the helicity states of the antitop quark correspond to spin projections 
 opposite to its momentum; namely, a positive (negative) helicity antitop has a 
 negative (positive) spin projection along its direction of motion.

In this basis, the density matrix can be written as
\begin{equation}
\rho =
\begin{pmatrix}
\rho_{+-,+-} & \rho_{+-,++} & \rho_{+-,--} & \rho_{+-,-+} \\
\rho_{++,+-} & \rho_{++,++} & \rho_{++,--} & \rho_{++,-+} \\
\rho_{--,+-} & \rho_{--,++} & \rho_{--,--} & \rho_{--,-+} \\
\rho_{-+,+-} & \rho_{-+,++} & \rho_{-+,--} & \rho_{-+,-+}
\end{pmatrix}.
\label{eq:rho-matrix-sec2}
\end{equation}

For brevity we denote these elements as
$\rho_{ij}$ $(i,j=1,\dots,4)$.
Since the density matrix is Hermitian, $\rho_{ij}=\rho_{ji}^{*}$,
it is convenient to express the polarization vectors and the
spin-correlation matrix in terms of the real and imaginary parts
of the matrix elements.
Using Eq.~\eqref{eq:BC-from-rho-sec2}, we obtain
\begin{align}
B_{1r} &= 2\,\mathrm{Re}(\rho_{13}+\rho_{24}), &
B_{1n} &= -2\,\mathrm{Im}(\rho_{13}+\rho_{24}), &
B_{1k} &= \rho_{11}+\rho_{22}-\rho_{33}-\rho_{44}, \\
B_{2r} &= 2\,\mathrm{Re}(\rho_{12}+\rho_{34}), &
B_{2n} &= -2\,\mathrm{Im}(\rho_{12}+\rho_{34}), &
B_{2k} &= \rho_{11}-\rho_{22}+\rho_{33}-\rho_{44},
\end{align}
and
\begin{align}
C_{rr} &= 2\,\mathrm{Re}(\rho_{14}+\rho_{23}), & 
C_{rn} &= 2\,\mathrm{Im}(\rho_{23}-\rho_{14}), & 
C_{rk} &= 2\,\mathrm{Re}(\rho_{13}-\rho_{24}), \label{eq:C1} \\
C_{nr} &= -2\,\mathrm{Im}(\rho_{14}+\rho_{23}), &
C_{nn} &= 2\,\mathrm{Re}(\rho_{23}-\rho_{14}), &
C_{nk} &= 2\,\mathrm{Im}(\rho_{24}-\rho_{13}), \\
C_{kr} &= 2\,\mathrm{Re}(\rho_{12}-\rho_{34}), &
C_{kn} &= 2\,\mathrm{Im}(\rho_{34}-\rho_{12}), &
C_{kk} &= \rho_{11}-\rho_{22}-\rho_{33}+\rho_{44}. \label{eq:C3}
\end{align}

In the following, the correlation coefficients $C_{ij}$ are computed explicitly
for each theoretical scenario considered in this work.
To this end, in Sec.~\ref{sec:amplitudes} we evaluate the helicity amplitudes for
 $l^-l^+ \to t\bar t$ in the SM and in the benchmark extensions,
and construct the spin matrix $R$ defined in Eq.~\eqref{eq:R-from-amps-sec2}.
From the resulting density matrix $\rho$, the coefficients $C_{ij}$ are
extracted using Eqs.~\eqref{eq:C1}-\eqref{eq:C3} and subsequently employed in
Sec.~\ref{sec:numerics} to evaluate the entanglement markers, the concurrence,
and the maximal CHSH parameter.

\section{Models}
\label{sec:models}

In this section we specify the three scenarios considered in the paper and collect the interactions 
relevant for $l^-l^+\to t\bar t$ $(l=e,\mu)$.

\subsection{Scalar mediator extension of the SM}
\label{subsec:scalar-model}

We first consider the SM extended by a $SU(2)$ doublet scalar field $\Phi$ with $U(1)_Y$ charge ${1/2}$ that
 couples to a charged lepton $l=e,~\mu$, and  to the top quark via Yukawa interactions.
The relevant Lagrangian terms are
\begin{equation}
  \mathcal{L} \supset
  - m_\Phi^2 \Phi^\dagger \Phi
  +\left\{- Y_l\, \bar{l}_L \Phi l_R
  - Y_t\, \bar{q}_L \tilde{\Phi} t_R+h.c.\right\},
  \label{eq:L-scalar}
\end{equation}
where $\tilde{\Phi}=i\sigma_2 \Phi^*$, 
 and $Y_l$ and $Y_t$ denote Yukawa couplings.
We assume that $\Phi$ develops no vacuum expectation value.

In this scenario, $l^-l^+\to t\bar t$ receives, in addition to the SM
$\gamma$ and $Z$ exchange, an $s$-channel scalar-exchange contribution
proportional to $Y_l Y_t$, which we will denote by $\mathcal{M}^{(\Phi)}$.
Its helicity structure differs from that of vector and axial-vector exchange,
and in particular allows same-helicity final states that are suppressed in
the pure SM.
As we will see, this feature has a direct impact on the entanglement pattern
of the final-state spins.

\subsection{Minimal gauged $U(1)_{B-L}$ model}
\label{subsec:BL-model}

The second scenario is the minimal gauged $U(1)_{B-L}$ extension of the 
 SM~\cite{Davidson:1978pm, Mohapatra:1980qe, Marshak:1979fm, Wetterich:1981bx, Masiero:1982fi, Mohapatra:1982xz, Buchmuller:1991ce}.
The gauge group is enlarged to
\begin{equation}
  SU(3)_C \times SU(2)_L \times U(1)_Y \times U(1)_{B-L},
\end{equation}
and the SM fermions carry $B-L$ charges equal to their baryon $(B)$ minus lepton $(L)$
number.
Anomaly cancellation requires the introduction of three right-handed neutrinos.
The $U(1)_{B-L}$ symmetry is spontaneously broken by a scalar field
with $U(1)_{B-L}$ charge $+2$, giving mass to the new gauge boson $Z'$ and generating
Majorana masses for the right-handed neutrinos.

The gauge-kinetic terms involving the two Abelian factors are
\begin{equation}
  \mathcal{L} \supset
  - \frac{1}{4} B_{\mu\nu} B^{\mu\nu}
  - \frac{1}{4} Z'_{\mu\nu} Z'^{\mu\nu}
  - \frac{\sin\chi}{2} B_{\mu\nu} Z'^{\mu\nu},
\end{equation}
where $B_\mu$ and $Z'_\mu$ are the gauge fields of $U(1)_Y$ and $U(1)_{B-L}$,
respectively, and $\chi$ parametrizes kinetic mixing.
After electroweak and $B-L$ symmetry breaking, the neutral gauge bosons mix
into the photon, the SM $Z$ and the heavy $Z'$ boson.
In the simplest limit of vanishing kinetic mixing at the $B-L$ breaking scale,
the mass eigenstate $Z'$ couples vectorially to fermions with strength
proportional to their $B-L$ charge.

For the purposes of this work, we assume $\chi=0$ in which the
relevant couplings of $Z'$ to leptons
and top quarks are
\begin{equation}
  \mathcal{L} \supset
  - g_{B-L} Q_{B-L}^{l}\, \bar l \gamma^\mu l\, Z'_\mu
  - g_{B-L} Q_{B-L}^t\, \bar t \gamma^\mu t\, Z'_\mu,
  \label{eq:L-Zprime}
\end{equation}
where $g_{B-L}$ denotes the gauge coupling constant of the $U(1)_{B-L}$ symmetry,
 and $Q_{B-L}^{l} = -1$ and $Q_{B-L}^t = 1/3$ are the corresponding $B\!-\!L$ charges.
The mass and the $U(1)_{B-L}$ coupling,
 $m_{Z'}$ and $g_{B-L}$, are treated as free parameters within the
 experimental bounds.

In this scenario, the $l^-l^+\to t\bar t$ amplitude receives a new $s$-channel
contribution from $Z'$ exchange, which interferes with the SM photon and
$Z$ amplitudes.
We denote this contribution by $\mathcal{M}^{(Z')}$.

\subsection{Randall-Sundrum model}
\label{subsec:RS-model}

As a third benchmark scenario, we consider the RS model with a single
warped extra dimension~\cite{Randall:1999ee}.
In this framework, the five-dimensional spacetime is a slice of
anti-de Sitter space compactified on an $S^1/\mathbb{Z}_2$ orbifold,
bounded by a Planck (UV) brane and a TeV (IR) brane.
The non-factorizable warped metric generates an exponential hierarchy
between the electroweak and Planck scales, thereby providing a geometrical
solution to the gauge hierarchy problem.

A characteristic feature of the RS model is that the zero-mode graviton and
its KK excitations have distinct wave-function profiles along
the extra dimension.
While the graviton zero mode is localized near the Planck brane and couples
to SM fields with the usual Planck-suppressed strength,
the KK gravitons are localized toward the TeV brane.
As a consequence, their couplings to SM fields residing on the TeV brane
are significantly enhanced and become relevant for collider phenomenology.

The effective interaction between the KK gravitons and the SM fields can be
written as~\cite{Davoudiasl:1999jd, Davoudiasl:2000wi}
\begin{equation}
  \mathcal{L}_{\rm int}
  = -\frac{1}{\bar{M}_{\rm Pl}}
  T^{\mu\nu}(x)\,h^{(0)}_{\mu\nu}(x)
  -\frac{1}{\Lambda_\pi}
  T^{\mu\nu}(x)\sum_{n=1}^{\infty} h^{(n)}_{\mu\nu}(x),
  \label{eq:RS-int}
\end{equation}
where $h^{(n)}_{\mu\nu}$ denotes the $n$-th KK graviton mode,
$T^{\mu\nu}$ is the energy-momentum tensor of the SM fields,
and $\Lambda_\pi=\bar{M}_{\rm Pl} e^{-\kappa r_c \pi}$ is an effective
interaction scale of order the TeV scale with $\bar{M}_{\rm Pl}=2.4\times 10^{18}~{\rm GeV}$
being the reduced Planck mass.
The contribution of the graviton zero mode is negligible for collider
processes, whereas each KK graviton couples to SM fields with strength
suppressed only by $\Lambda_\pi$.

The masses of the KK gravitons are given by
\begin{equation}
  m_n = x_n\,\kappa\, e^{-\kappa r_c \pi},
\end{equation}
where $x_n$ are the roots of the Bessel function $J_1(x_n)=0$,
with $x_1\simeq 3.83$, $x_2\simeq 7.02$, $x_3\simeq 10.17$, etc.
Once the mass of the lightest KK graviton $m_1$ is fixed,
the entire KK spectrum is determined by $m_n=m_1(x_n/x_1)$.
It is convenient to trade the parameters of the RS geometry for the
phenomenological input parameters $m_1$ and $\kappa/\bar{M}_{\rm Pl}$,
with theoretical consistency requiring the five-dimensional curvature $\kappa$ to be
$\kappa/\bar{M}_{\rm Pl}\ll 1$,
typically $\kappa/\bar{M}_{\rm Pl}\lesssim 0.1$~\cite{Davoudiasl:2000wi}.

In the RS scenario, the process $l^-l^+\to t\bar t$ receives additional
contributions from $s$-channel exchange of virtual KK gravitons.
The corresponding invariant amplitude can be written schematically as
\begin{equation}
  \mathcal{M}^{(G)} =
  A(s)\, T^{\rm in}_{\mu\nu}(p_1,p_2)\eta^{\mu\rho}\eta^{\nu\sigma}\, T^{\rm out}_{\rho\sigma}(k_1,k_2),
\end{equation}
with $p_1$ and $p_2$ ($k_1$ and $k_2$) denoting the momenta of the initial (final) particles, and
\begin{equation}
  A(s) = -\frac{1}{\Lambda_\pi^2}
  \sum_{n=1}^{\infty}
  \frac{1}{s - m_n^2 + i m_n \Gamma_n},\quad s=(p_1+p_2)^2=(k_1+k_2)^2, \label{eq:A}
\end{equation}
where $\Gamma_n$ denotes the total decay width of the $n$-th KK graviton given by 
\begin{eqnarray}
 \Gamma_n(h^{(n)}\rightarrow yy)
 ={m_n x_n^2 \over 16 \pi}\left(
 {\kappa \over \bar{M}_{\rm pl}}\right)^2
 \sum_{y}\Delta_n^{yy}\,,
\end{eqnarray}
with $\Delta_n^{yy}$ being a dimensionless coefficient 
for each decay mode, whose explicit forms are given by
\begin{align}
\Delta_n^{\gamma\gamma} &= \frac{1}{5}, \quad
\Delta_n^{gg} = \frac{8}{5}, \quad
\Delta_n^{\nu\bar{\nu}} = \frac{1}{10}, \\
\Delta_n^{VV} &= c_V \,\sqrt{1-4r_V}
\left(\frac{13}{12}+\frac{14}{3}r_V+4r_V^2\right),
\quad (V = W,Z), \\
\Delta_n^{HH} &= \frac{1}{30}(1-4r_H)^{5/2}, \\
\Delta_n^{f\bar{f}} &= c_f \,(1-4r_f)^{3/2}
\left(1+\frac{8}{3}r_f\right),
\quad (f = l,q),
\end{align}
with
\begin{equation}
c_W = \frac{2}{5}, \quad
c_Z = \frac{1}{5}, \quad
c_l = \frac{1}{10}, \quad
c_q = \frac{3}{10},
\end{equation}
and $r_y = m_y^2/m_n^2$ for each SM particle $y$.
Expressions for leptons and quarks are for one flavor. 

The energy momentum tensors are given by
\begin{align}
T^{\text{in}}_{\mu\nu}
&= \bar{v}(p_2,\lambda_{l^+})\,
V^{\text{in}}_{\mu\nu}\,u(p_1,\lambda_{l^-}),
\\
T^{\text{out}}_{\mu\nu}
&= \bar{u}(k_1,\lambda_t)\,
V^{\text{out}}_{\mu\nu}\,
v(k_2, \lambda_{\bar{t}}).
\end{align}
with
\begin{align}
V^{\text{in}}_{\mu\nu}
&= \frac{1}{4} \left[
(p_1 - p_2)_\mu \gamma_\nu
+ (p_1 - p_2)_\nu \gamma_\mu
+ \eta_{\mu\nu}
\left(
-\frac{1}{2} (p_1 - p_2)_\rho \gamma^\rho
\right)
\right],
\\[6pt]
V^{\text{out}}_{\mu\nu}
&= \frac{1}{4} \left[
(k_1 - k_2)_\mu \gamma_\nu
+ (k_1 - k_2)_\nu \gamma_\mu
+ \eta_{\mu\nu}
\left(
-\frac{1}{2} (k_1 - k_2)_\rho \gamma^\rho +m_t\right)
\right],
\end{align}
where $\gamma^\mu$ are the gamma matrices, $m_t$ denotes the top-quark mass, 
 $\eta_{\mu\nu} = \mathrm{diag}(1,-1,-1,-1)$ is the Minkowski metric, 
 and $u(p,\lambda)$ and $v(p,\lambda)$ denote helicity spinors 
 given in Eqs.~(\ref{eq:sp1})--(\ref{eq:sp4}) in Sec.~\ref{subsec:kinematics}.
 
The exchange of spin-2 KK gravitons induces characteristic angular
 distributions and helicity structures that are qualitatively different
 from those arising from scalar or vector mediators.
As a result, the RS model can lead to distinctive modifications of the
 spin correlations and quantum-entanglement properties of the produced
 $t\bar t$ system.
In our analysis, we include KK modes up to $i=4$, since higher modes
 develop large decay widths and cease to appear as well-defined resonances,
 making their individual contributions difficult to identify.

\section{Helicity amplitude and density matrix for $l^-l^+ \to t\bar t$}
\label{sec:amplitudes}
In this section, we present the helicity amplitudes for $l^-l^+\to t\bar t$ $(l=e,\mu)$
in the SM and in the new-physics scenarios. The construction of the spin
density matrix and quantum-information observables has been defined in
Sec.~\ref{sec:formalism}.

\subsection{Kinematics and spinor conventions}
\label{subsec:kinematics}

We work in the center-of-mass frame of the incoming leptons, with 4-momenta
\begin{align}
  p_{1}^\mu &= \frac{\sqrt{s}}{2} (1, 0, 0, 1), \\
  p_{2}^\mu &= \frac{\sqrt{s}}{2} (1, 0, 0, -1), \\
  k_1^\mu     &= \frac{\sqrt{s}}{2} (1, \beta \sin\theta\cos\phi, \beta\sin\theta\sin\phi, \beta \cos\theta), \\
  k_{2}^\mu
              &= \frac{\sqrt{s}}{2} (1, -\beta \sin\theta\cos\phi, -\beta\sin\theta\sin\phi, -\beta \cos\theta),
\end{align}
where
\begin{equation}
  \beta = \sqrt{1 - \frac{4 m_t^2}{s}}
\end{equation}
is the top-quark velocity in the center-of-mass frame.
The angle $\theta$ is the scattering angle of the top quark with respect to 
the $l^-$ beam, 
taken along the $z$ axis. 
The azimuthal angle $\phi$ is measured around the beam axis from the $x$ axis, so that the scattering plane is specified by the plane spanned by 
the incoming 
lepton
momentum and the top-quark momentum.

We use spinors in the helicity basis for a fermion with mass $m$ and four-momentum
 $p^\mu = (E, \bm{p})$, where $E = \sqrt{|\bm{p}|^2 + m^2}$, given by
\begin{align}
  u(p,+1/2) &=
  \begin{pmatrix}
    -\sqrt{E-|\bf p|}\,\chi(\hat{\bm{p}}) \\
    -\sqrt{E+|\bf p|}\,\chi(\hat{\bm{p}})
  \end{pmatrix}, \label{eq:sp1} \\
  u(p,-1/2) &=
  \begin{pmatrix}
    \sqrt{E+|\bf p|}\,\xi(\hat{\bm{p}}) \\
    \sqrt{E-|\bf p|}\,\xi(\hat{\bm{p}})
  \end{pmatrix}, \\
  v(p,+1/2) &=
  \begin{pmatrix}
    -\sqrt{E+|\bf p|}\,\xi(\hat{\bm{p}}) \\
    \sqrt{E-|\bf p|}\,\xi(\hat{\bm{p}})
  \end{pmatrix}, \\
  v(p,-1/2) &=
  \begin{pmatrix}
    \sqrt{E-|\bf p|}\,\chi(\hat{\bm{p}}) \\
    -\sqrt{E+|\bf p|}\,\chi(\hat{\bm{p}})
  \end{pmatrix}, \label{eq:sp4}
\end{align}
where $\hat{\bm{p}}=\bm{p}/|\bm{p}|$ is the direction of the three-momentum.
The two-component helicity spinors are
\begin{equation}
  \chi(\hat{\bm{p}}) =
  \begin{pmatrix}
    \cos\frac{\theta}{2} \\
    e^{i\phi}\sin\frac{\theta}{2}
  \end{pmatrix},
  \quad
  \xi(\hat{\bm{p}}) =
  \begin{pmatrix}
    -e^{-i\phi}\sin\frac{\theta}{2} \\
    \cos\frac{\theta}{2}
  \end{pmatrix},
\end{equation}
with $(\theta,\phi)$ the polar and azimuthal angles of $\hat{\bm p}$. 
For the present calculation, we choose the scattering plane to coincide with the $x$-$z$ plane.
Accordingly, the initial 
lepton and antilepton
are taken along the $+z$ and $-z$ directions, corresponding to $(\theta,\phi)=(0,0)$ and $(\pi,0)$, respectively.
The final-state top quark is assigned $(\theta,\phi)=(\theta,0)$, while the anti-top quark is assigned $(\pi-\theta,\pi)$.
This choice entails no loss of generality, since the azimuthal dependence is absorbed into the definition of the orthonormal basis $(\hat r,\hat n,\hat k)$ introduced in Eq.~(\ref{helbasis}). Consequently, all spin-dependent observables depend only on the polar angle $\theta$.

Our convention for the fermion spinors is based on the HELAS convention~\cite{Murayama:1992gi}.
However, it differs from the HELAS convention by an overall sign in the right-handed particle spinors.
This difference originates from a convention-dependent phase choice in the definition of the spinors.
In the present work, we adopt a convention in which this sign is explicitly retained.
With this choice, the resulting density matrix satisfies the relation, 
\begin{equation}
R_{\lambda_t\lambda_{\bar t},\lambda'_t\lambda'_{\bar t}}
=
R_{\lambda_{\bar t}\lambda_t,\lambda'_{\bar t}\lambda'_t}.
\end{equation}
if the interaction is CP invariant \cite{Bernreuther:1993hq}.
Although the interactions considered in this work are not CP invariant, 
 we adopt this convention so that the above relation holds in the CP-conserving limit, 
 such as in QCD.

\subsection{SM, scalar, $Z'$ and graviton contributions to the amplitude}
\label{subsec:amps}

The full helicity amplitude can be written as
\begin{equation}
  \mathcal{M}({\lambda_{l^-}, \lambda_{l^+}; \lambda_t, \lambda_{\bar t}})
  = \mathcal{M}^{\rm (SM)}({\lambda_{l^-}, \lambda_{l^+}; \lambda_t, \lambda_{\bar t}})
  + \mathcal{M}^{(X)}({\lambda_{l^-}, \lambda_{l^+}; \lambda_t, \lambda_{\bar t}}).
  \label{eq:amp-decomp}
\end{equation}
As introduced above, we denote the helicities $-1/2$ and $+1/2$ by $-$ and $+$, respectively.
Here $\mathcal{M}^{(\rm SM)}$ denotes the SM contribution, while
 $\mathcal{M}^{(X)}$ represents the contribution from physics beyond the SM,
 with $X=\Phi$ for the scalar mediator, $X=Z'$ for the $U(1)_{B-L}$ model,
 and $X=G$ for the RS scenario, respectively.
 
The first term in the right-hand side is of SM, which is given by
\begin{align}
\mathcal{M}^{(\rm SM)}(+,-;\pm, \pm)&= \sqrt{1-\beta^{2}}\,\sin\theta\,
\left[e^{2} Q_{l} Q_{t}+{s\, g_{Z}^{Rl} \over 2}\frac{g_{Z}^{Lt}+g_{Z}^{Rt}}{s - m_{Z}^{2}+im_Z\Gamma_Z}\right], \\
\mathcal{M}^{(\rm SM)}(-,+;\pm,\pm)&=-\sqrt{1-\beta^{2}}\,\sin\theta\,
\left[e^{2} Q_{l} Q_{t}+{s\, g_{Z}^{Ll} \over 2}\frac{g_{Z}^{Lt}+g_{Z}^{Rt}}{s - m_{Z}^{2}+im_Z\Gamma_Z}\right], \\
\mathcal{M}^{(\rm SM)}(+,-;\pm,\mp) &= 
 \left(1\pm \cos\theta\right)\left[e^{2} Q_{l} Q_{t} +{s\, g_{Z}^{Rl} \over 2}\frac{\,g_{Z}^{Lt}(1\mp\beta)+g_{Z}^{Rt}(1\pm\beta)}{s - m_{Z}^{2}+im_Z\Gamma_Z}\right], \nonumber \\ \\
\mathcal{M}^{(\rm SM)}(-,+;\pm,\mp)&=(1\mp \cos\theta)\left[\,e^{2} Q_{l} Q_{t}
+{s\, g_{Z}^{Ll} \over 2}\frac{g_{Z}^{Lt}(1\mp\beta) + g_{Z}^{Rt}(1 \pm\beta)}{s - m_{Z}^{2}+im_Z\Gamma_Z}\right],  \nonumber \\
\end{align}
where $m_Z$ and $\Gamma_{Z}$ is the mass and the decay width of $Z$ boson, and
\begin{eqnarray}
  &\displaystyle Q_{l} =-1, \quad Q_t = \frac{2}{3}, & \\
  & \displaystyle g_Z^{La}={e \over \cos\theta_W \sin\theta_W}\left(T_3^a-Q_a \sin^2\theta_W \right), \quad
    g_Z^{Ra}=-{e Q_a \sin\theta_W \over \cos\theta_W}, &
\end{eqnarray}
with $a=l, t$ and
\begin{eqnarray}
  \displaystyle T_3^{l}=-{1 \over 2},\quad T_3^t={1 \over 2}.
\end{eqnarray}

The additional contributions from the scalar are as follows:
\begin{align}
\mathcal{M}^{(\rm \Phi)}(\pm,\pm;\pm,\pm)&= -{s \over 2}\frac{\, Y_{l} Y_{t}\,(1-\beta)}{s - m_{\Phi}^{2}+im_{\Phi}\Gamma_{\Phi}}, \\
\mathcal{M}^{(\rm \Phi)}(\pm,\pm;\mp,\mp)&= {s \over 2}\frac{\, Y_{l} Y_{t}\,(1+\beta)}{s - m_{\Phi}^{2}+im_{\Phi}\Gamma_{\Phi}},
  \label{eq:amp-scalar}
\end{align}
For the benchmark choice $m_{\Phi}=200~\mathrm{GeV}$ adopted in our numerical analysis, 
the channel $\Phi\to t\bar t$ is closed kinematically, and we assume no other relevant decay modes. 
Hence $\Phi$ decays only into $l^-l^+$ $(l=e,\mu)$, so that
\begin{equation}
\Gamma_{\Phi}\simeq \Gamma(\Phi\to l^-l^+)
= \frac{|Y_l|^2\,m_{\Phi}}{16\pi}.
\end{equation}

For the $U(1)_{B-L}$ scenario, we have
\begin{align}
  \mathcal{M}^{(Z')}(\pm,\mp; +,+)&=\pm \sqrt{1-\beta^{2}}\sin\theta\frac{s\, g_{B-L}^{2}Q^{l}_{B-L}Q^t_{B-L}}{s - m_{Z'}^{2}+im_{Z'}\Gamma_{Z'}}, \\
  \mathcal{M}^{(Z')}(\pm,\mp; -,-)&=\pm \sqrt{1-\beta^{2}} \sin\theta\frac{s\, g_{B-L}^{2}Q^{l}_{B-L}Q^t_{B-L}}{s - m_{Z'}^{2}+im_{Z'}\Gamma_{Z'}}, \\
  \mathcal{M}^{(Z')}(\pm,\mp;+,-)&=(1\pm\cos\theta)\frac{s\, g_{B-L}^{2}Q^{l}_{B-L}Q^t_{B-L}}{s - m_{Z'}^{2}+im_{Z'}\Gamma_{Z'}}, \\
  \mathcal{M}^{(Z')}(\pm,\mp;-,+)&= (1\mp\cos\theta)\frac{s\, g_{B-L}^{2}Q^{l}_{B-L}Q^t_{B-L}}{s - m_{Z'}^{2}+im_{Z'}\Gamma_{Z'}},
   \label{eq:amp-Zprime}
\end{align}
where $m_{Z^\prime}$ is the mass of $Z^\prime$ boson and $\Gamma_{Z'}$ is the decay width of $Z'$ given by $\Gamma_{Z'}={13 \over 24\pi}g_{B-L}^2m_{Z'}$.
This has the same Lorentz structure as photon exchange but with different
 couplings and a massive propagator.

For the RS scenario, the additional contribution arises from
$s$-channel exchange of KK gravitons.
The corresponding helicity amplitudes can be written as
\begin{align}
\mathcal{M}^{(G)}(\pm,\mp; +,+)
 &= \mp {s^2 \beta \over 4}A(s) \,\sqrt{1-\beta^{2}}\,\sin\theta\cos\theta, \label{eq:amp-graviton1}\\
\mathcal{M}^{(G)}(\pm,\mp; -,-)
 &= \mp {s^2\beta \over 4}  A(s)\,\sqrt{1-\beta^{2}}\,\sin\theta\cos\theta, \label{eq:amp-graviton2}\\
\mathcal{M}^{(G)}(\pm,\mp; +, -)
 &= \mp {s^2\beta \over 8} A(s)\,(2\cos^2\theta \pm \cos\theta - 1), \label{eq:amp-graviton3} \\
\mathcal{M}^{(G)}(\pm,\mp; -, +)
 &= \pm{s^2\beta \over 8} A(s)\,(2\cos^2\theta \mp \cos\theta -1). \label{eq:amp-graviton4}
\end{align}
Unlike the scalar and vector mediators, the exchange of a spin-2 KK graviton
 induces helicity structures characteristic of its tensorial coupling to the
 energy-momentum tensor. 
As seen from Eqs.~\eqref{eq:amp-graviton1}--\eqref{eq:amp-graviton4},
 the angular dependence exhibits nontrivial combinations such as
 $\cos^2\theta$ and $\sin\theta\cos\theta$, which are absent in scalar or
 vector exchanges. 

Moreover, the overall energy dependence encoded in $A(s)$ differs
 significantly from that of scalar or vector mediators, effectively scaling as $s$ at the amplitude level. 
This distinct behavior reflects the
 higher-dimensional origin of the interaction and can lead to qualitatively
 different modifications of spin correlations and entanglement observables,
 especially at high energies.
 
While the $U(1)_{B-L}$ contribution has the same Lorentz structure as photon
 exchange, differing only by the couplings and the massive propagator,
 the RS contribution represents a genuinely new tensor interaction.
As a consequence, the RS scenario provides a qualitatively distinct benchmark
 for studying the sensitivity of quantum-entanglement observables to the spin
 and Lorentz structure of new physics.

\section{Numerical analysis}
\label{sec:numerics}
In this section we present numerical results for the entanglement marker,
 the concurrence, and the maximal CHSH Bell parameter in the four scenarios.
The SM inputs are fixed to $\alpha=e^2/(4\pi)=1/128$, $m_t=173~\mathrm{GeV}$,
 $\sin^2\theta_W=0.23343$ and $m_Z=91.188~\mathrm{GeV}$.
For the numerical analysis, we adopt representative benchmark points
 for each new-physics scenario, as summarized in Table~\ref{Table:benchmark}.
The parameters are chosen within phenomenologically viable ranges,
 and the new-physics contributions lead to visible deviations
 from the SM.
Our goal is not merely to quantify the amount of quantum correlation in each 
 benchmark model. 
Rather, we investigate whether quantum-information observables can reveal characteristic 
signatures of new particles and the structure of their interactions. 
To this end, we compare scalar, vector, and tensor mediators on equal footing and examine 
 how the distinctive features of these new particles and their interactions are encoded in the 
 kinematic dependence of the entanglement marker, concurrence, and Bell-nonlocality observables.  

\begin{table}[b]
	\center
\begin{tabular}{cccc}
 \hline	
	model & mediated particle & masses [TeV] & coupling parameters \\ \hline
    scalar-mediator & $\Phi$ (spin-0) & 0.2 & $Y_l=Y_t=0.2$ \\
	 $U(1)_{B-L}$ & $Z'$ (spin-1) & 5.0 & $g_{B-L}=0.4$ \\
	 RS & $G$ (spin-2) & 5.5, 10.1, 14.6, 19.1 & $\kappa/\bar{M}_{\mathrm{Pl}}=0.2$ \\	 
 \hline	
\end{tabular}
\caption{Benchmark masses and coupling parameters in the new-physics models for numerical analysis.}	
\label{Table:benchmark}
\end{table}

\begin{figure}[htbp]
\begin{center}
  \begin{minipage}[b]{0.45\linewidth}
    \centering
    \includegraphics[keepaspectratio, scale=0.29]{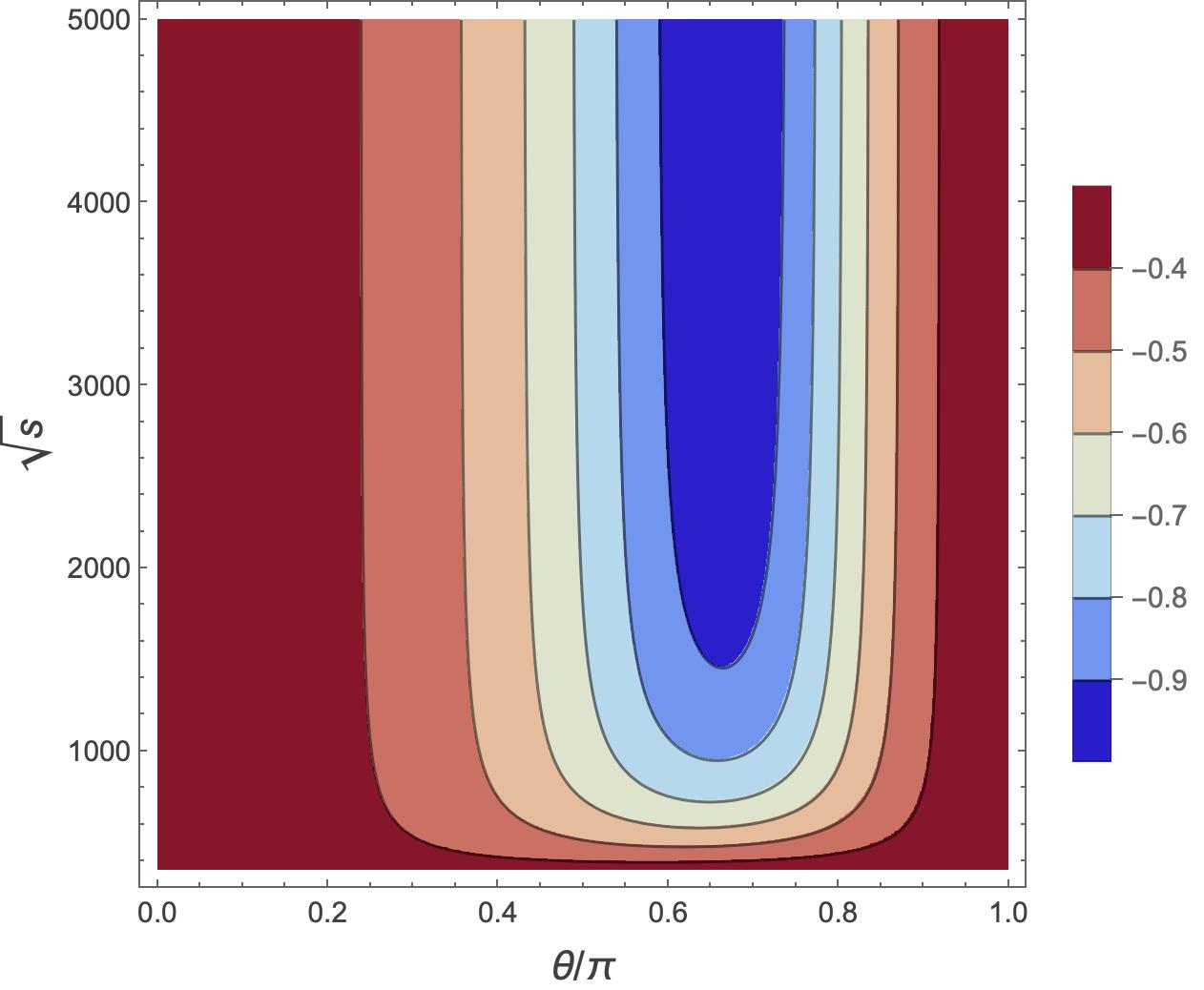}
  \end{minipage}
  \begin{minipage}[b]{0.45\linewidth}
    \centering
    \includegraphics[keepaspectratio, scale=0.29]{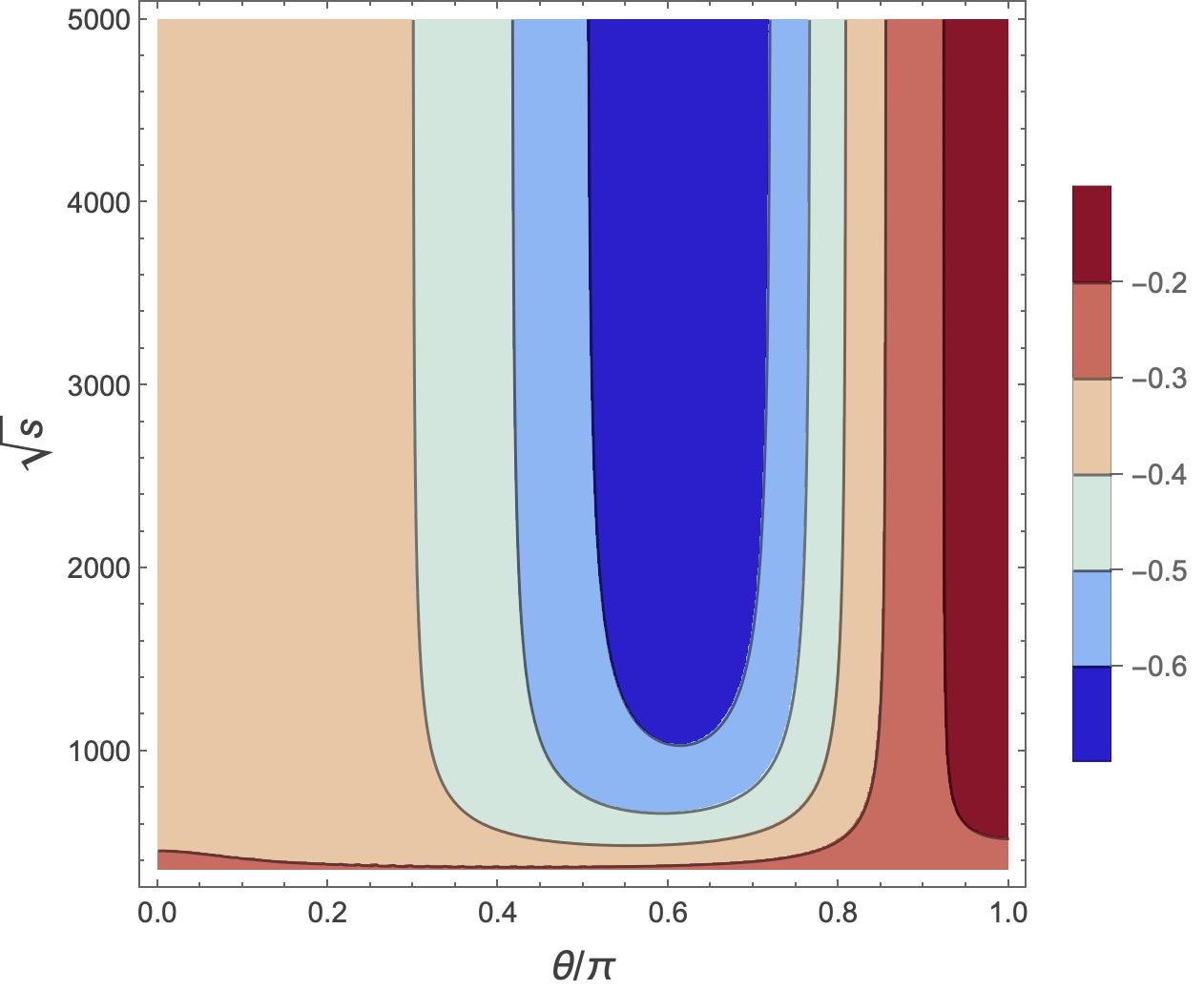}
  \end{minipage} \par\medskip
  \begin{minipage}[b]{0.45\linewidth}
    \centering
    \includegraphics[keepaspectratio, scale=0.29]{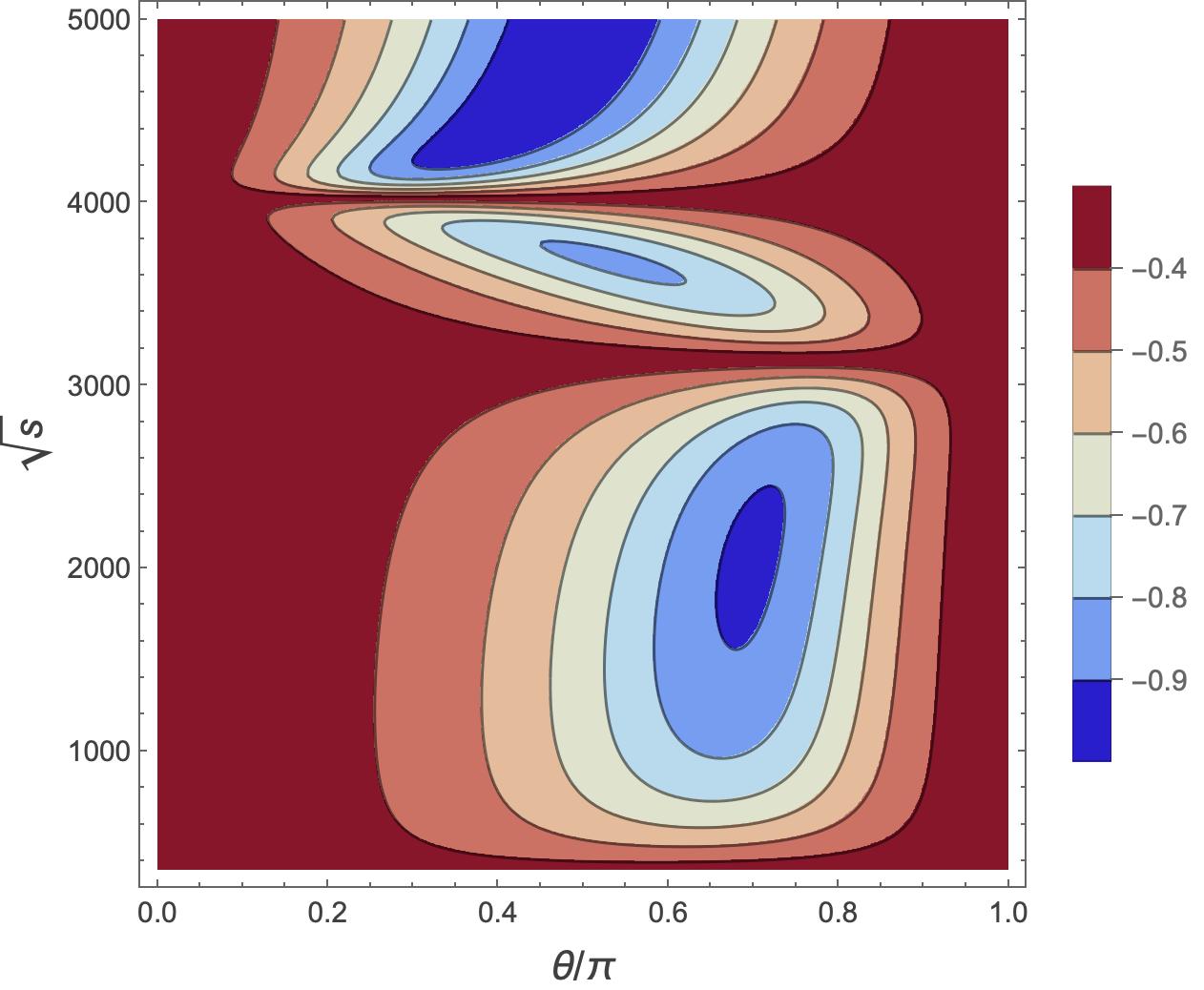}
  \end{minipage}
  \begin{minipage}[b]{0.45\linewidth}
    \centering
    \includegraphics[keepaspectratio, scale=0.29]{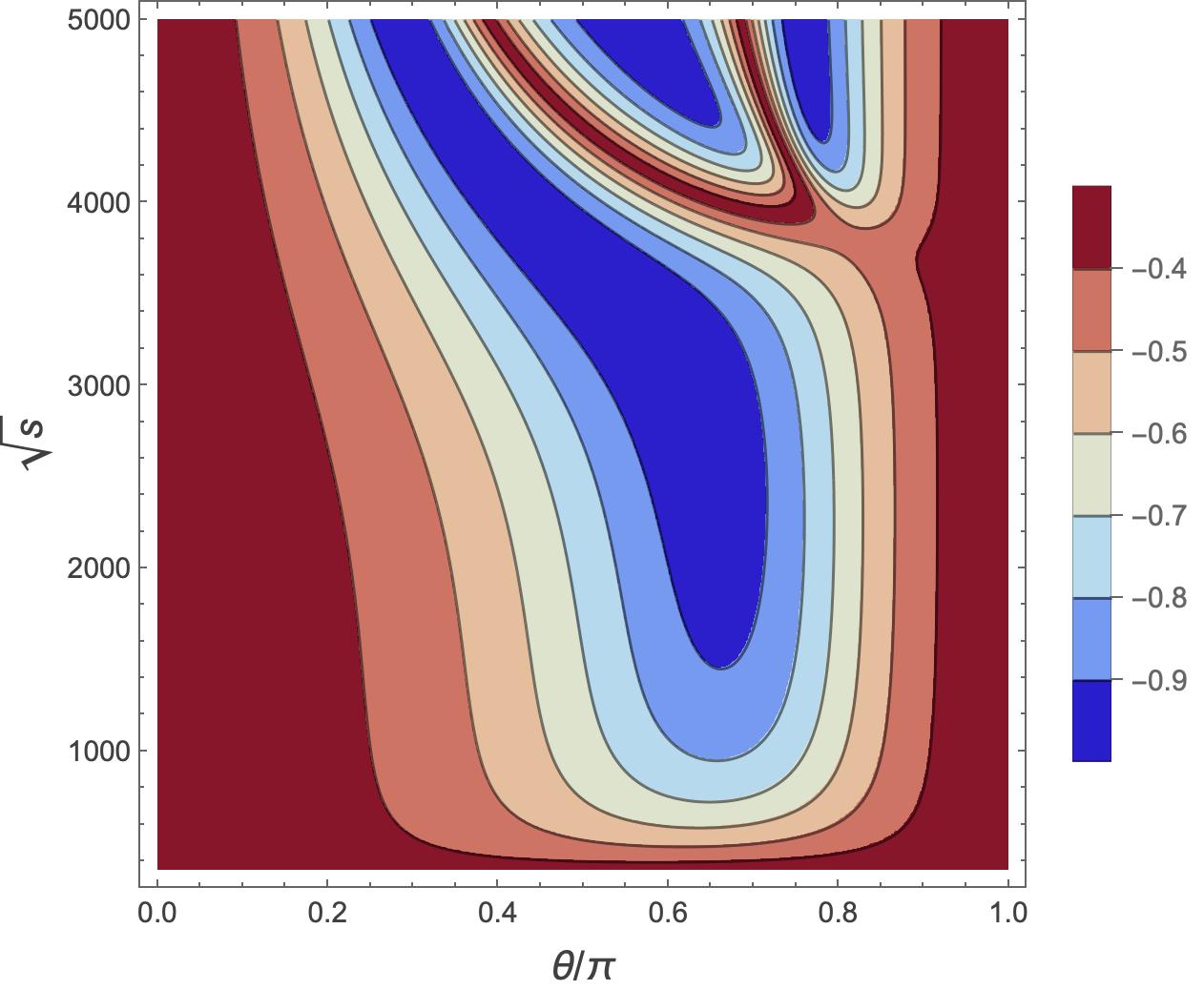}
  \end{minipage}
\caption{Contours of the entanglement marker in the $(\sqrt{s},\theta/\pi)$ plane.
The top panels show the SM (left) and the scalar-mediator model (right),
while the bottom panels show the $U(1)_{B-L}$ model (left) and the RS model (right).
}
\label{fig:EmarkerContours_sqrts}
\end{center}
\end{figure}

\begin{figure}[htbp]
\begin{center}
  \begin{minipage}[b]{0.45\linewidth}
    \centering
    \includegraphics[keepaspectratio, scale=0.3]{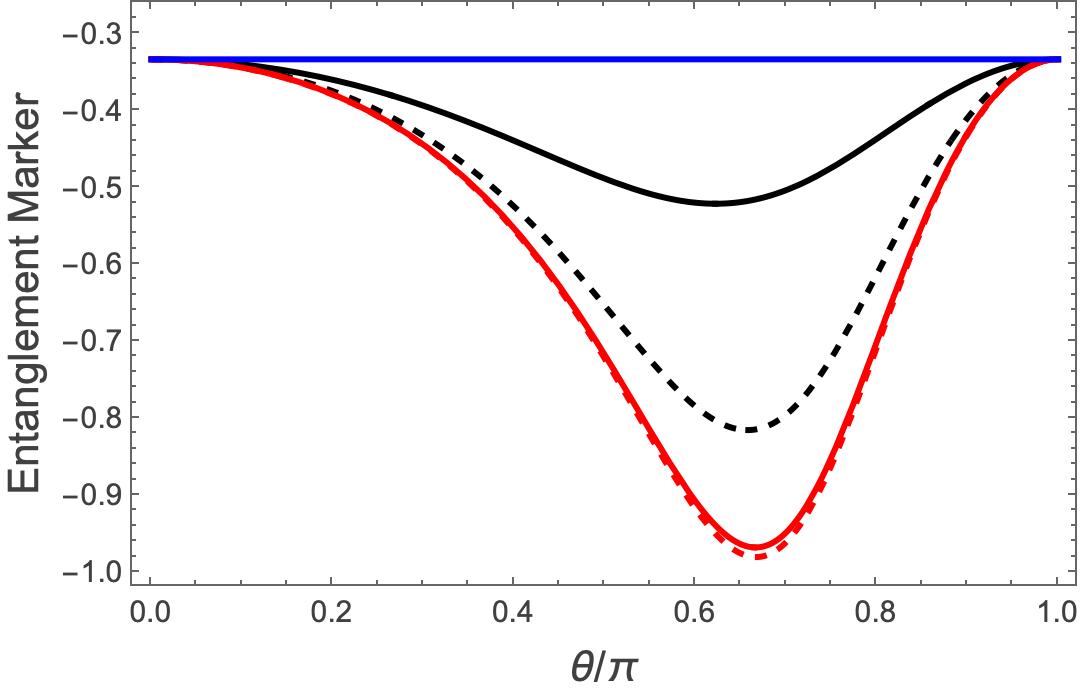}
  \end{minipage}
  \begin{minipage}[b]{0.45\linewidth}
    \centering
    \includegraphics[keepaspectratio, scale=0.3]{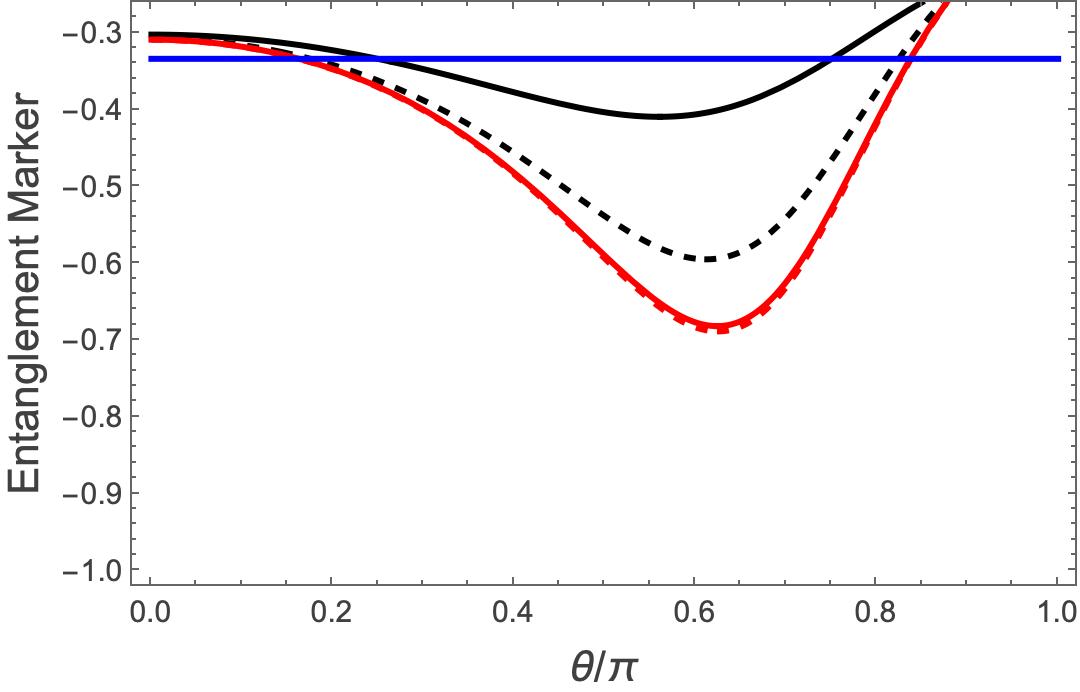}
  \end{minipage} \par\medskip
  \begin{minipage}[b]{0.45\linewidth}
    \centering
    \includegraphics[keepaspectratio, scale=0.3]{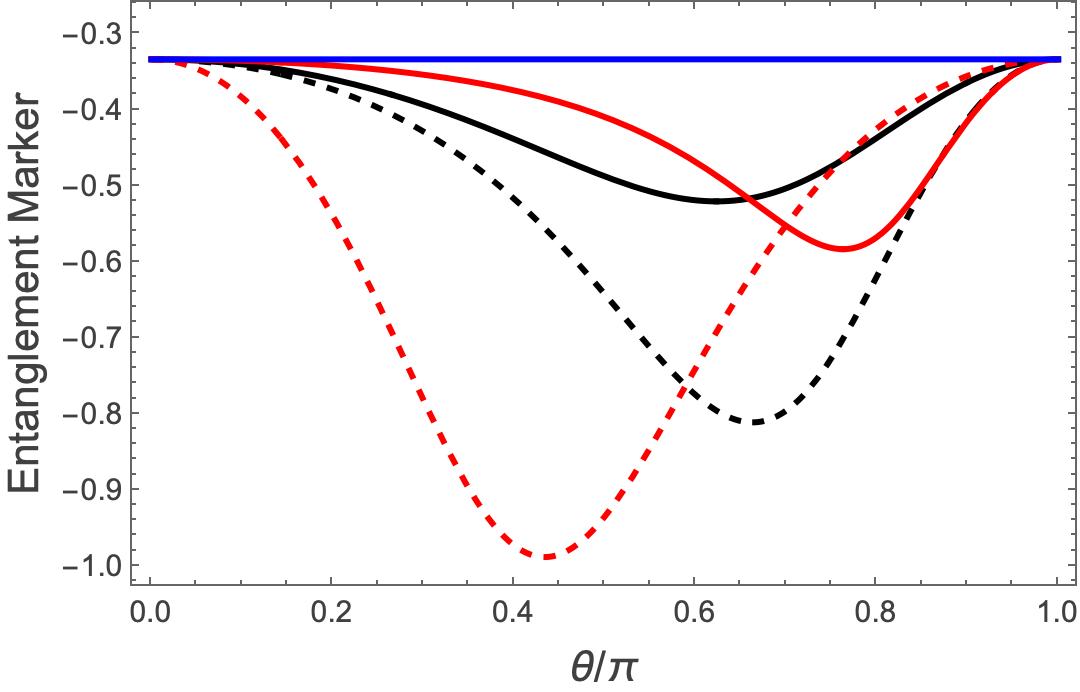}
  \end{minipage}
  \begin{minipage}[b]{0.45\linewidth}
    \centering
    \includegraphics[keepaspectratio, scale=0.3]{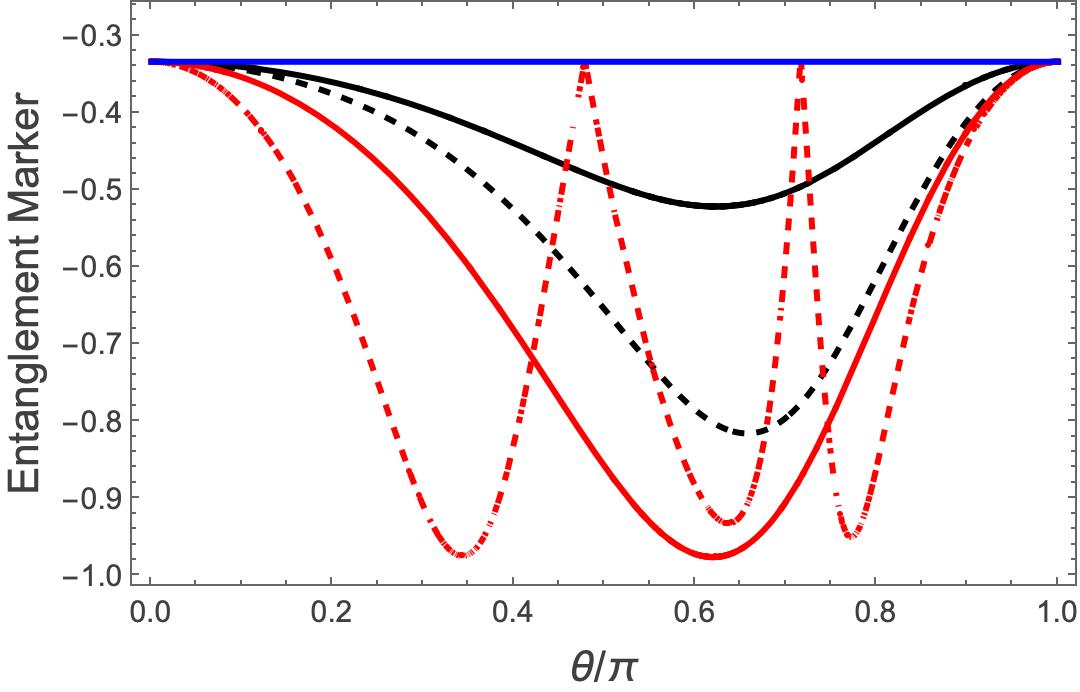}
  \end{minipage}
\caption{Angular dependence of the entanglement marker at fixed center-of-mass energies
 for the SM (top left), the scalar-mediator model (top right),
 the $U(1)_{B-L}$ model  (bottom left) and RS model.
The solid black, dashed black, solid red, and dashed red curves correspond to
$\sqrt{s}=500~\mathrm{GeV}$, $1000~\mathrm{GeV}$, $3000~\mathrm{GeV}$ and $4500~\mathrm{GeV}$, respectively.
The horizontal blue line indicates the threshold below which the state is
entangled.
}
\label{fig:EmarkerContoursSlices}
\end{center}
\end{figure}

We first discuss the entanglement marker, shown in
Fig.~\ref{fig:EmarkerContours_sqrts} and Fig.~\ref{fig:EmarkerContoursSlices}.
The contour plots reveal the kinematic regions where quantum correlations are enhanced,
 while the angular distributions at fixed energies provide a complementary one-dimensional view
 that makes the location and depth of the extrema more transparent.

The top-left panel in Fig.~\ref{fig:EmarkerContours_sqrts} shows the SM case.
The entanglement marker exhibits a single connected structure,
with a smooth dependence on both the scattering angle and the center-of-mass energy.
The correlation is strongest in the central region,
$\theta/\pi \sim 0.5$--$0.7$, and increases gradually with energy.
This behavior is also evident in the corresponding energy slices
shown in the top-left panel of Fig.~\ref{fig:EmarkerContoursSlices},
where a single broad minimum develops and deepens as $\sqrt{s}$ increases.

The scalar mediator scenario, shown in the top-right panels
of Figs.~\ref{fig:EmarkerContours_sqrts} and \ref{fig:EmarkerContoursSlices},
preserves the overall angular pattern of the SM, with the entanglement marker
developing a broad minimum around intermediate scattering angles.
Its magnitude is, however, reduced.
This reduction can be understood from the helicity structure of the scalar interaction. 
Unlike the SM vector interactions, scalar exchange enhances like-helicity top-pair configurations, 
 which dilute the opposite-helicity spin correlations responsible for the strong quantum correlations 
 in the SM. 
Since the scalar amplitude does not interfere with the SM contribution in the massless-lepton limit, 
 the overall topology of the distribution remains largely unchanged and no qualitatively 
 new structures emerge.
As the center-of-mass energy increases, the minimum becomes deeper while its position
remains nearly unchanged. Narrow angular regions then emerge where the marker exceeds
the entanglement threshold, indicating a loss of entanglement.

In contrast, the $U(1)_{B-L}$ model leads to a more intricate structure,
 with multiple disconnected regions appearing in the contour plots,
 as shown in the bottom-left panel of Fig.~\ref{fig:EmarkerContours_sqrts}.
At low energies, the contour resembles the SM pattern.
At intermediate energies, a distinct entanglement extremum appears
between $\sqrt{s}\simeq3~{\rm TeV}$ and $4~{\rm TeV}$, while the
entanglement is significantly weakened near these two energies.
This produces the isolated entangled region observed in the contour plot.
At higher energies, the entanglement extremum appears around
$\theta/\pi=1/2$.
The analytic interpretation of these features is presented in
Appendices~\ref{app:chiral_vector} and
\ref{app:vector_interference}.
Appendix~\ref{app:chiral_vector} considers a generic chiral gauge
interaction and explains the angular position of the entanglement
extremum at low and high energies in terms of the relative strengths
of the left- and right-chiral couplings.
Appendix~\ref{app:vector_interference} extends this analysis by
including both a chiral SM-like interaction and a vector-like
interaction.
It shows that the energy dependence of the effective chiral
combinations defined in (\ref{eq:PQR_energy_dependent1}) and (\ref{eq:PQR_energy_dependent2}) 
governs the motion of the entanglement extremum in the
intermediate-energy region, while the weakened-entanglement regions
around $\sqrt{s}\simeq3~{\rm TeV}$ and $4~{\rm TeV}$ arise from the
cancellation condition, Eq.~(\ref{eq:interference_zero_condition}),
which suppresses the off-diagonal element of the density matrix.

In the corresponding energy slices shown in the bottom-left panel of
Fig.~\ref{fig:EmarkerContoursSlices}, the angular minimum moves with
$\sqrt{s}$, while its depth is strongly reduced around
$\sqrt{s}\simeq3~{\rm TeV}$, in agreement with the analytic
interpretation presented in Appendices~\ref{app:chiral_vector} and
\ref{app:vector_interference}.

The RS scenario, shown in the bottom-right panels of
Fig.~\ref{fig:EmarkerContours_sqrts},
exhibits a qualitatively different behavior 
from both the scalar and $U(1)_{B-L}$ scenarios.
Unlike these models, the RS contribution originates from spin-2 KK gravitons,
 which contribute through tensor interactions.
At relatively low energies, the contour plot exhibits a single connected
 region, similar to the SM behavior.
As the collision energy increases, however, additional disconnected
 regions emerge, indicating the growing impact of the KK-graviton
 contribution.
To clarify the origin of the disconnected angular regions,
 we analyze in Appendix~\ref{app:RS_angular}
 the contribution from the first KK graviton alone in the
 massless-top limit.
The analysis shows that the disconnected regions arise solely from the
 characteristic angular dependence of the spin-2 exchange.
The resulting angular distribution is symmetric about $\theta/\pi=1/2$
 as shown in Fig. \ref{fig:RS_first_KK}.
While the disconnected regions originate from the characteristic
 angular dependence of the spin-2 exchange, the full RS prediction
 contains an additional feature.
The contours are noticeably asymmetric about $\theta/\pi=1/2$.
This asymmetry is absent for the KK-graviton contribution alone but
 arises from interference with the SM amplitudes, particularly the
 $Z$-boson exchange.
Such features are absent in both the scalar and vector scenarios,
 leading to a quantum-correlation pattern that is qualitatively distinct
 from those found in the other benchmark models.

The same features are reflected in the corresponding energy slices
shown in the bottom-right panel of Fig.~\ref{fig:EmarkerContoursSlices}.
At relatively low energies, the angular distribution exhibits a single
broad minimum, similar to the SM behavior.
As the energy increases, additional local minima develop in the angular
distribution.
These minima correspond to the disconnected angular regions observed in
the contour plots and are consistent with the spin-2 angular structure
discussed in Appendix~\ref{app:RS_angular}.
Their positions are further modified by interference with the SM
amplitudes.
The angular distributions also exhibit the same asymmetry about
 $\theta/\pi=1/2$ observed in the contour plots, consistent with the interference 
  between the KK-graviton and SM contributions.

\begin{figure}[htbp]
\begin{center}
  \begin{minipage}[b]{0.45\linewidth}
    \centering
    \includegraphics[keepaspectratio, scale=0.3]{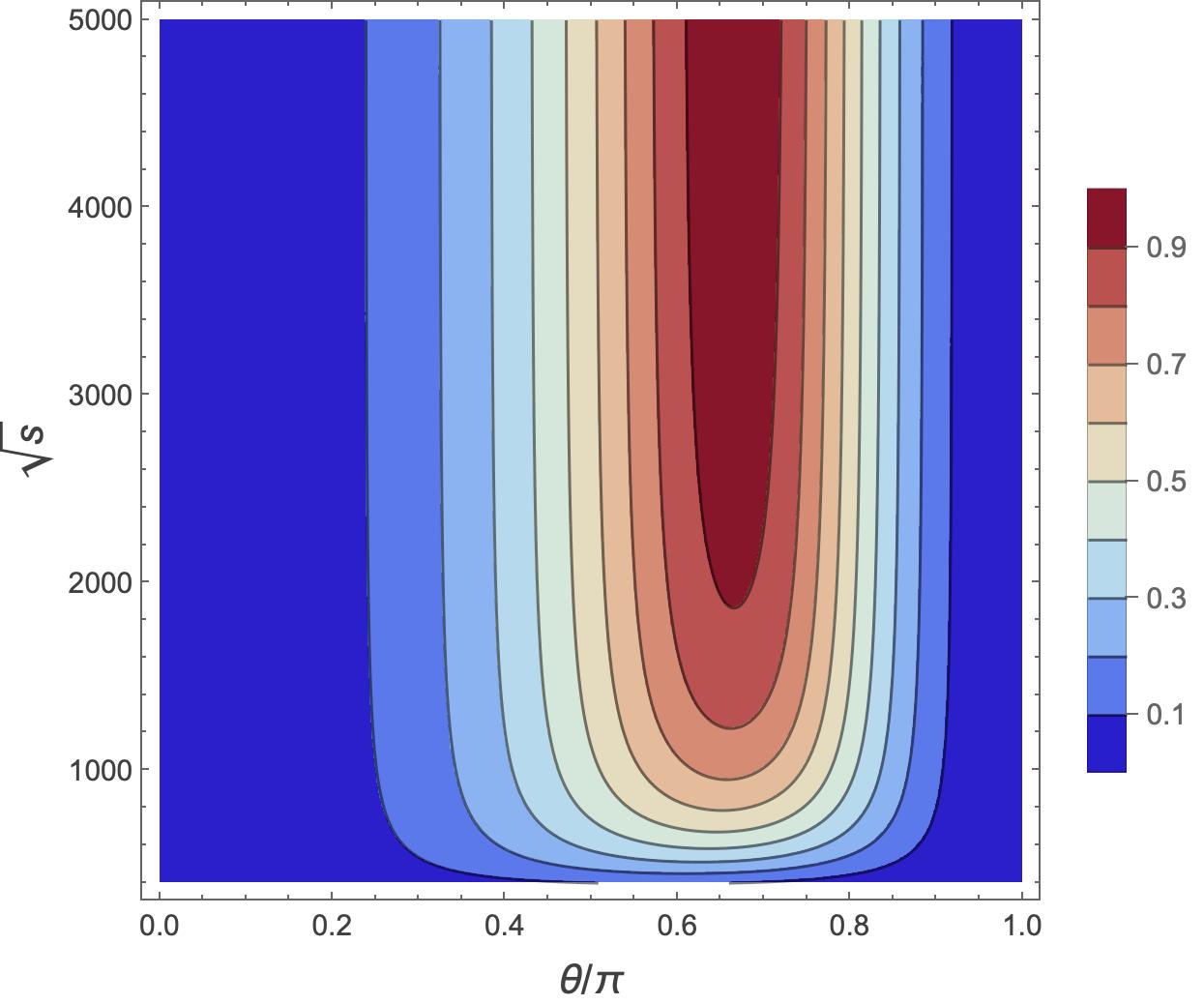}
  \end{minipage}
  \begin{minipage}[b]{0.45\linewidth}
    \centering
    \includegraphics[keepaspectratio, scale=0.3]{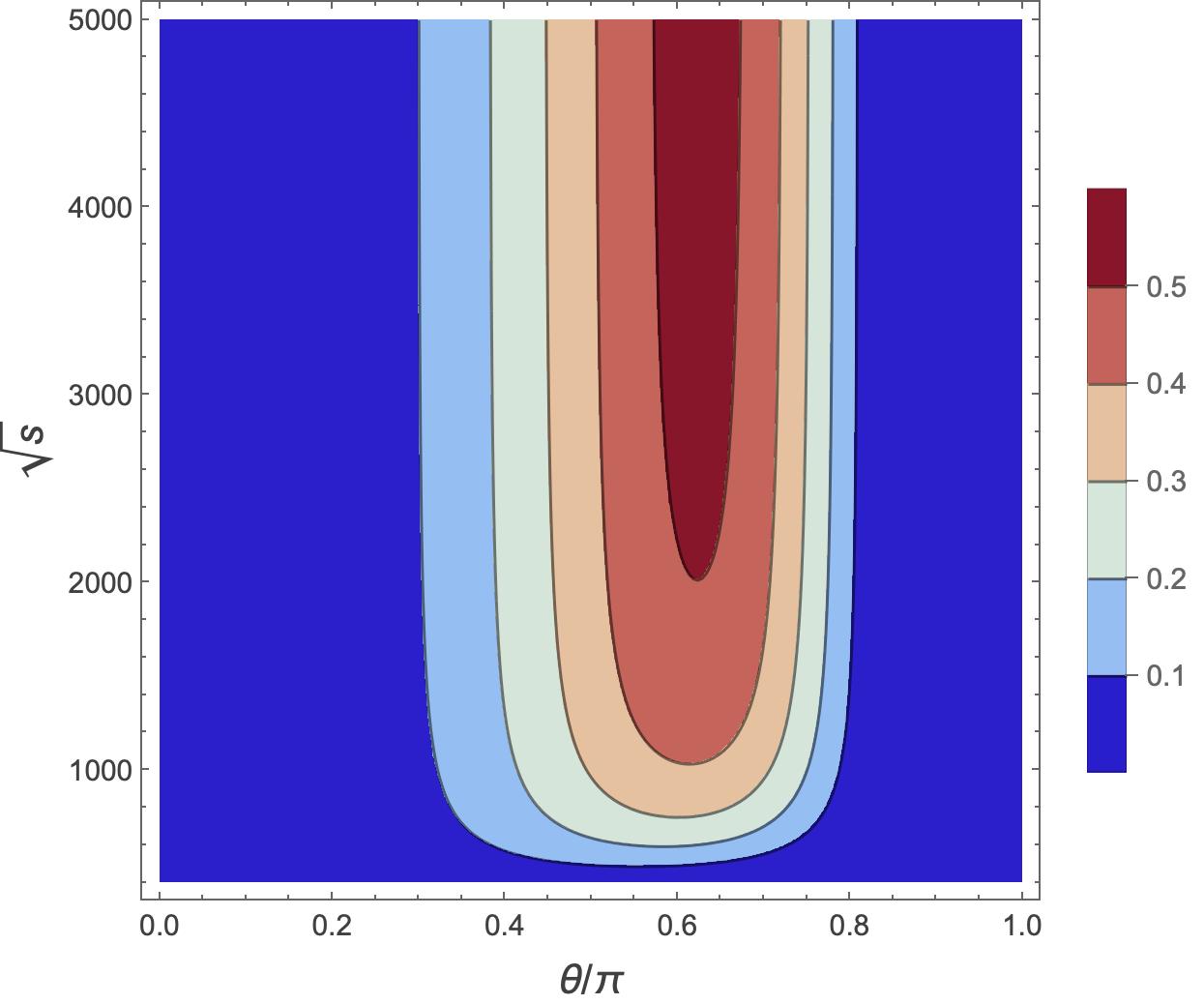}
  \end{minipage} \par\medskip
  \begin{minipage}[b]{0.45\linewidth}
    \centering
    \includegraphics[keepaspectratio, scale=0.3]{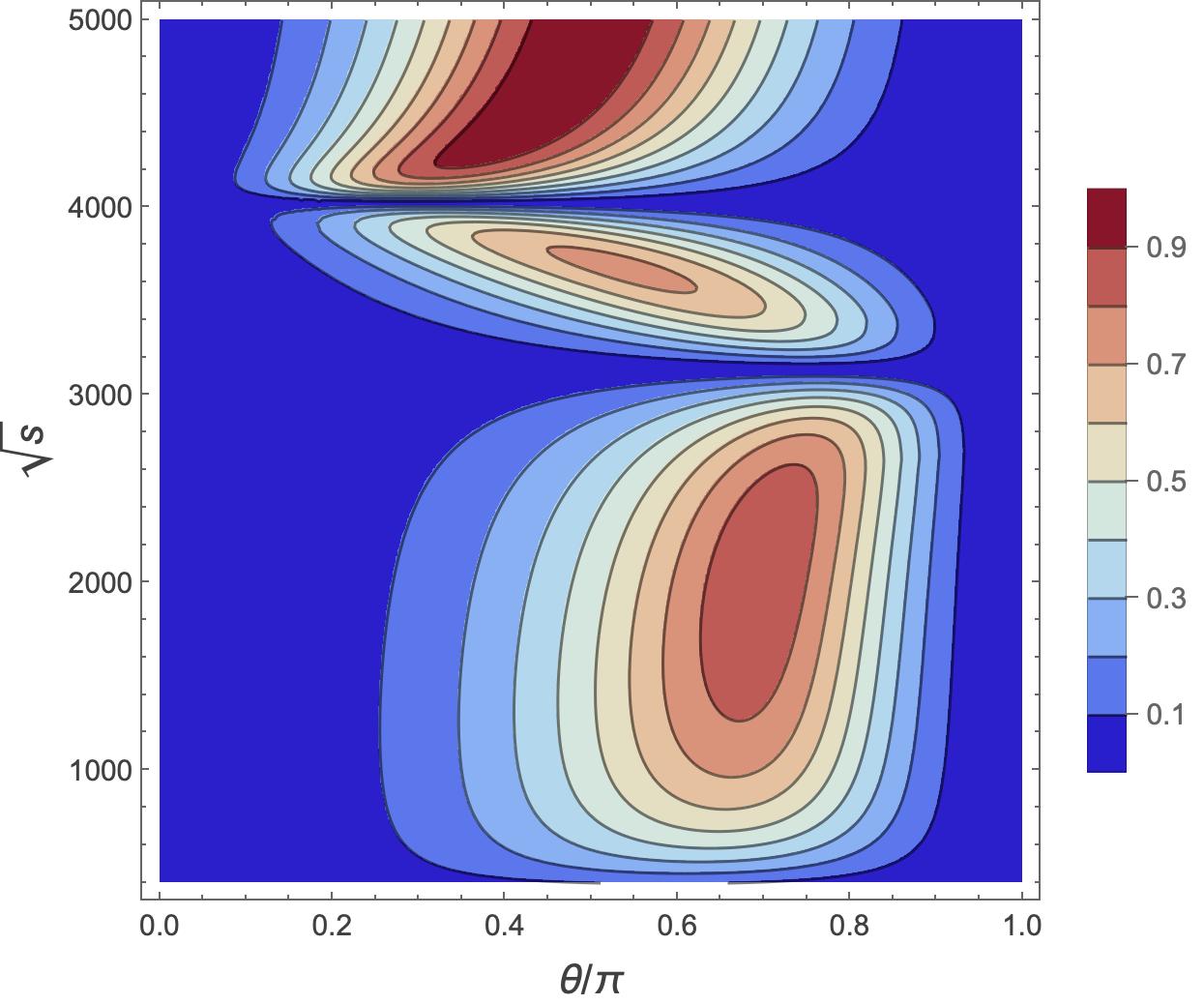}
  \end{minipage}
  \begin{minipage}[b]{0.45\linewidth}
    \centering
    \includegraphics[keepaspectratio, scale=0.3]{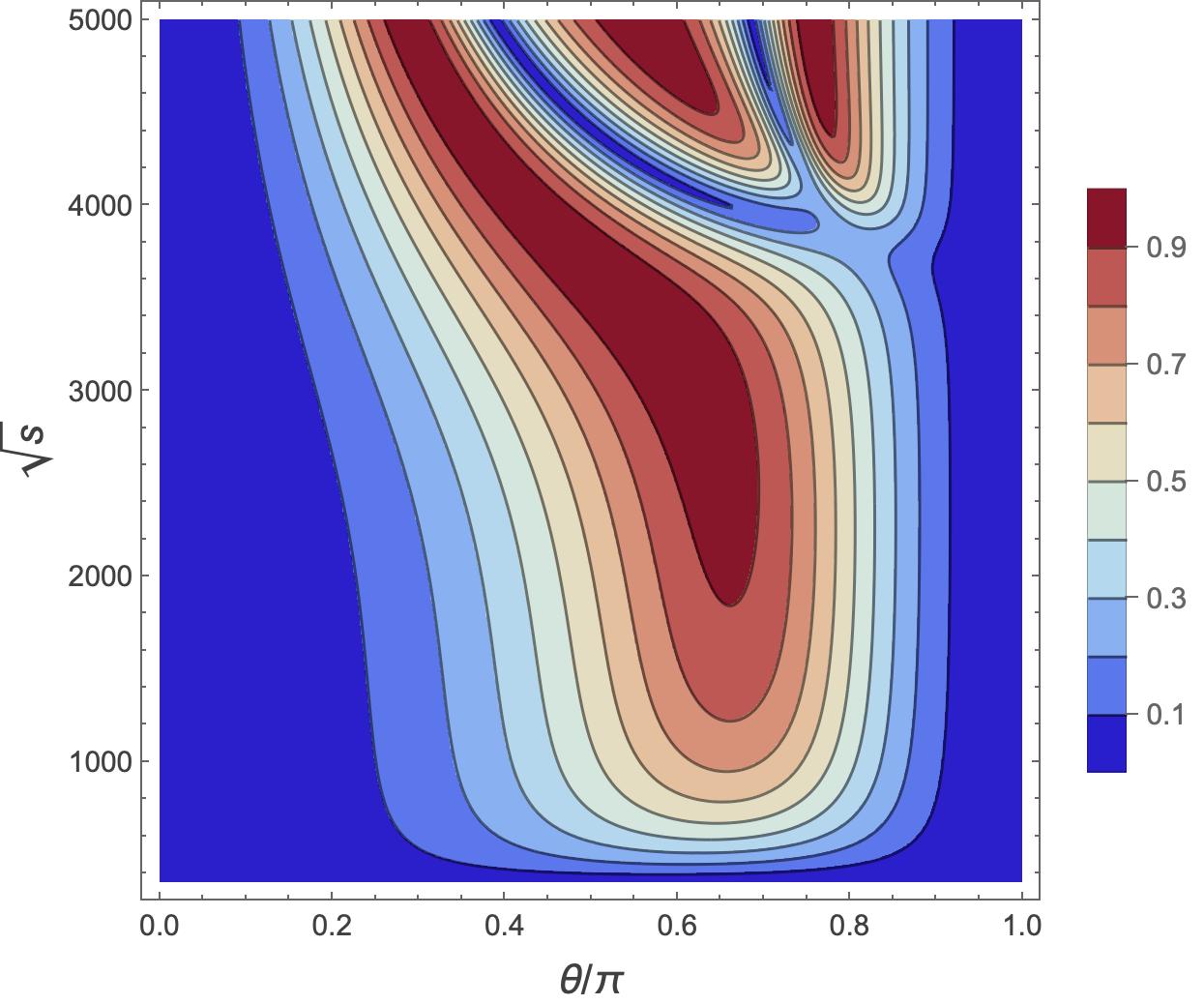}
  \end{minipage}
\caption{Contours of the concurrence in the $(\sqrt{s},\theta/\pi)$ plane.
The panel assignments are the same as in Fig.~\ref{fig:EmarkerContours_sqrts}
 for the entanglement marker.}
\label{fig:ConcurrenceContours_sqrts}
\end{center}
\end{figure}

\begin{figure}[htbp]
\begin{center}
  \begin{minipage}[b]{0.45\linewidth}
    \centering
    \includegraphics[keepaspectratio, scale=0.3]{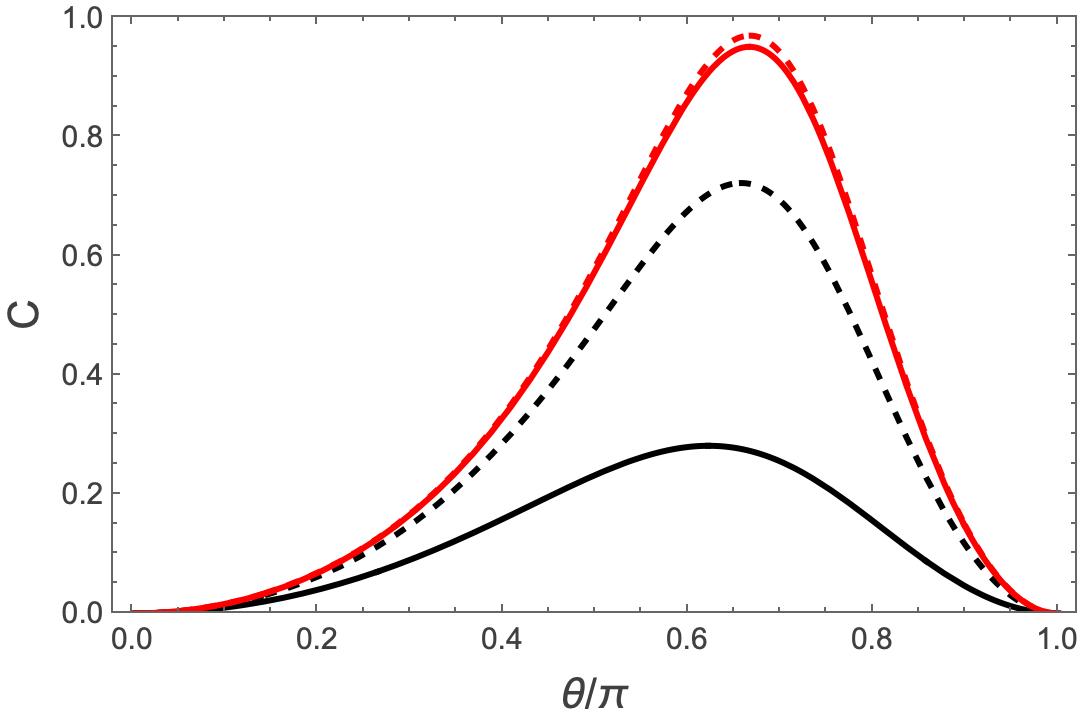}
  \end{minipage}
  \begin{minipage}[b]{0.45\linewidth}
    \centering
    \includegraphics[keepaspectratio, scale=0.3]{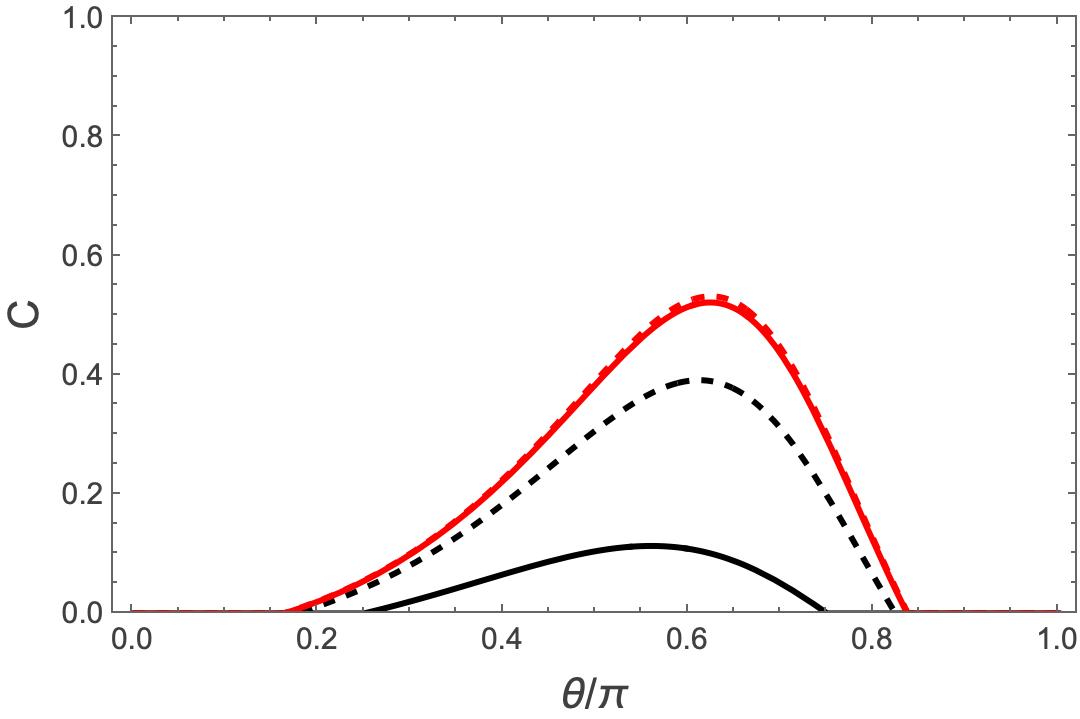}
  \end{minipage} \par\medskip
  \begin{minipage}[b]{0.45\linewidth}
    \centering
    \includegraphics[keepaspectratio, scale=0.3]{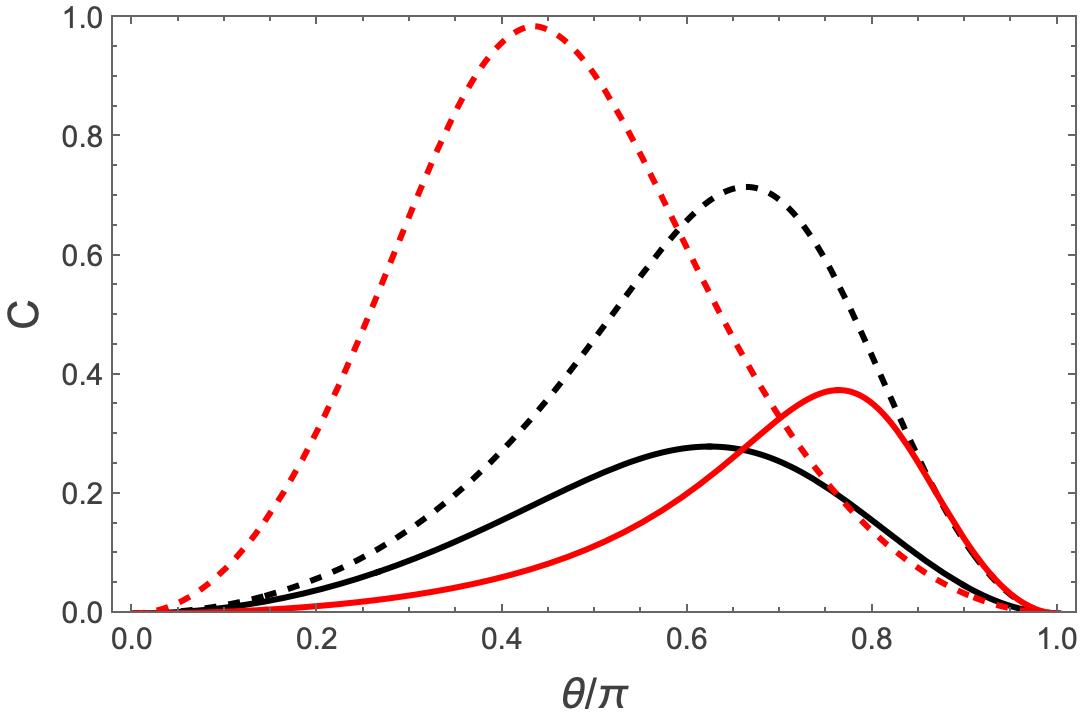}
  \end{minipage}
  \begin{minipage}[b]{0.45\linewidth}
    \centering
    \includegraphics[keepaspectratio, scale=0.3]{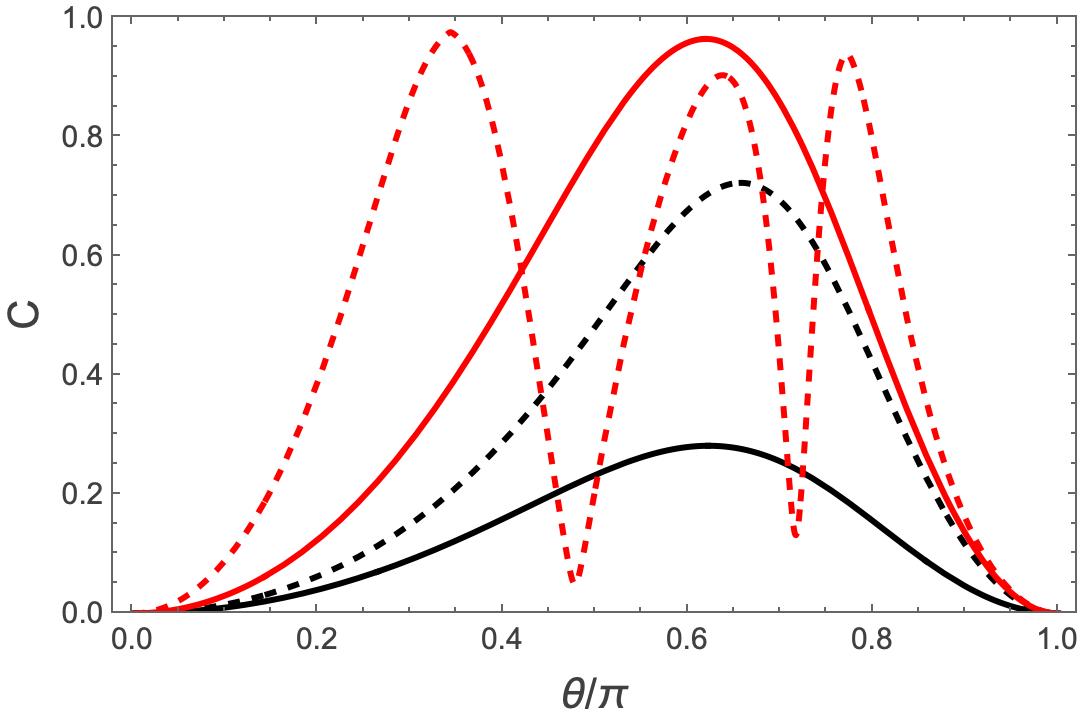}
  \end{minipage}
\caption{Angular dependence of the concurrence at fixed center-of-mass energies.
The panel assignments are the same as in Fig.~\ref{fig:EmarkerContoursSlices}.
The solid black, dashed black, solid red, and dashed red curves correspond to
$\sqrt{s}=500~\mathrm{GeV}$, $1000~\mathrm{GeV}$, $3000~\mathrm{GeV}$, and $4500~\mathrm{GeV}$, respectively.}
\label{fig:ConcurrenceContoursSlices}
\end{center}
\end{figure}

We now turn to the concurrence.
It is instructive to examine whether the qualitative differences among
 the scalar, vector, and RS scenarios observed in the entanglement marker
 persist in another entanglement observable.
The concurrence are shown in
 Fig.~\ref{fig:ConcurrenceContours_sqrts} and Fig.~\ref{fig:ConcurrenceContoursSlices}.
Overall, its behavior closely follows that of the entanglement marker,
 since both quantify the strength of quantum correlations.
In particular, the locations of the extrema and the global structure
 in the contour plots are nearly identical.
This correspondence is also reflected in the energy slices,
 where the angular dependence exhibits a similar structure,
 with the positions of the extrema largely aligned with those
 of the entanglement marker.

The main difference lies in the overall normalization,
 as the concurrence provides a more direct measure of entanglement.
The SM and scalar cases again show a smooth single-region structure,
 with a single maximum in the angular distribution, as shown in the top panels
 of Fig.~\ref{fig:ConcurrenceContoursSlices}.
The $U(1)_{B-L}$ model, shown in the bottom-left panel of
Fig.~\ref{fig:ConcurrenceContoursSlices},
retains a single maximum whose position evolves with
$\sqrt{s}$.
The maximum is strongly suppressed at
$\sqrt{s}=3~{\rm TeV}$ and increases again at higher energies.
In contrast, the RS scenario, shown in the bottom-right panel of
 Fig.~\ref{fig:ConcurrenceContoursSlices}, develops multiple local maxima,
 confirming the qualitatively distinct pattern already observed in the
 entanglement marker.
These observations confirm the qualitative picture obtained
 from the entanglement marker.

\begin{figure}[htbp]
\begin{center}
  \begin{minipage}[b]{0.45\linewidth}
    \centering
    \includegraphics[keepaspectratio, scale=0.3]{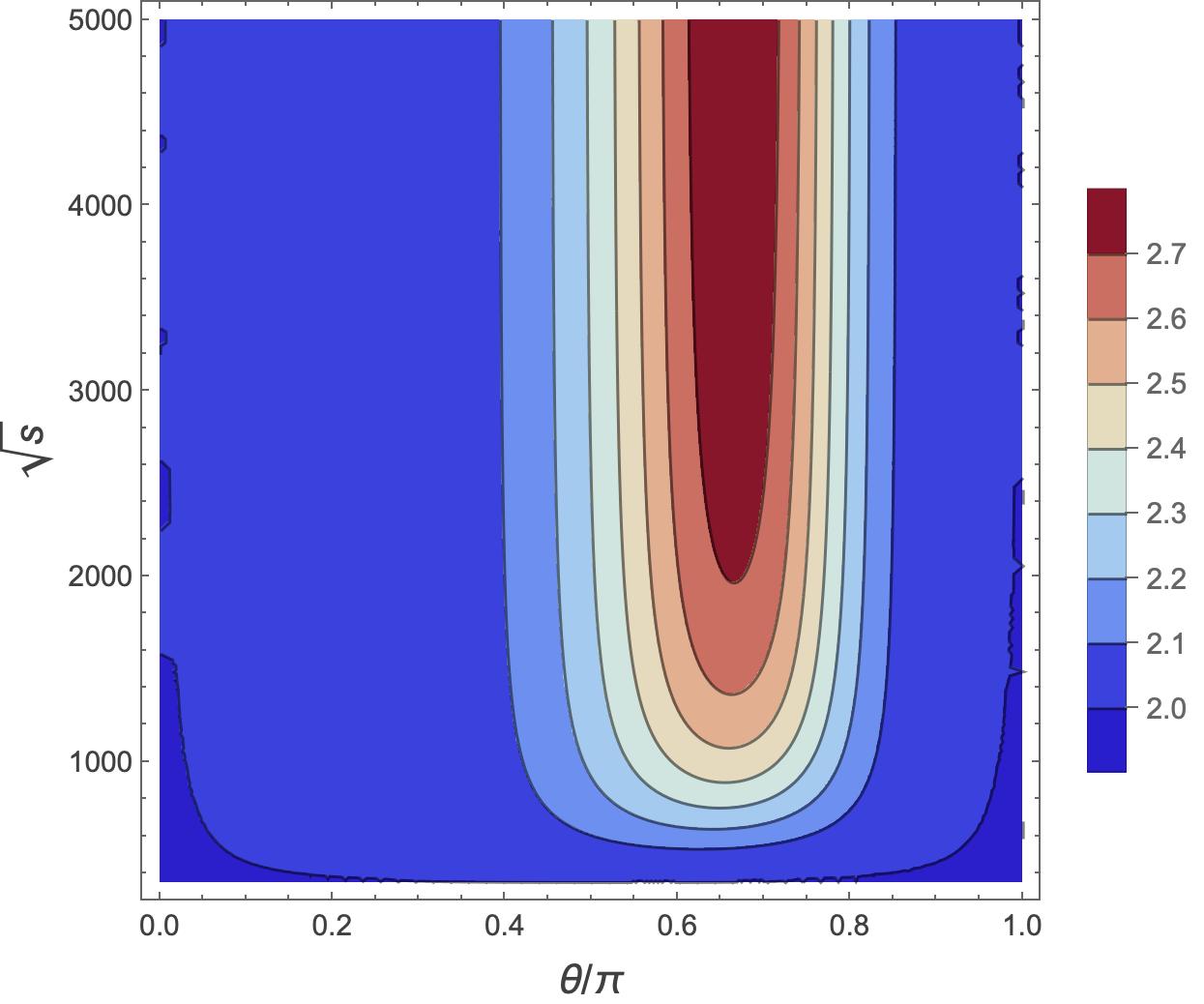}
  \end{minipage}
  \begin{minipage}[b]{0.45\linewidth}
    \centering
    \includegraphics[keepaspectratio, scale=0.3]{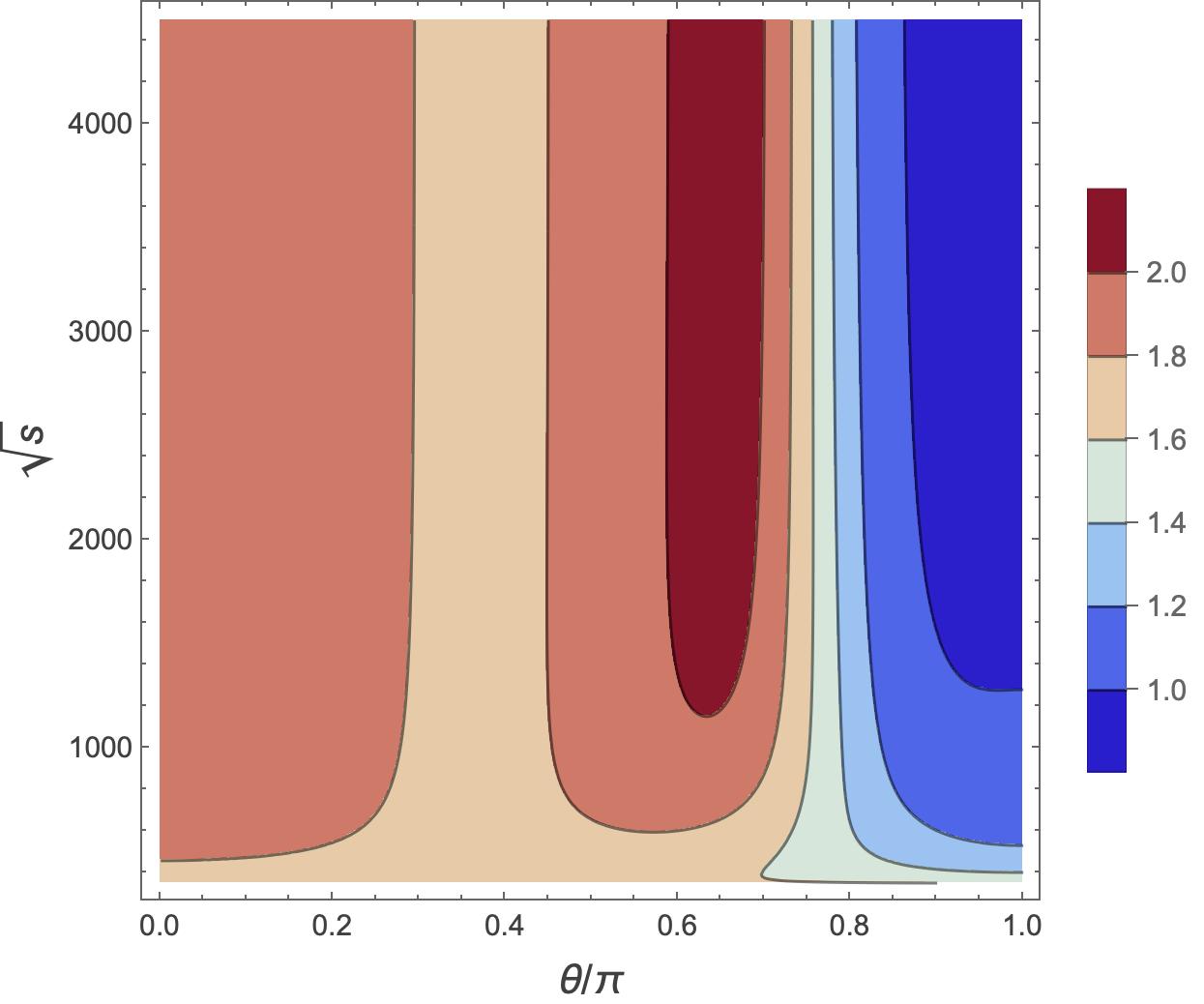}
  \end{minipage} \par\medskip
  \begin{minipage}[b]{0.45\linewidth}
    \centering
    \includegraphics[keepaspectratio, scale=0.3]{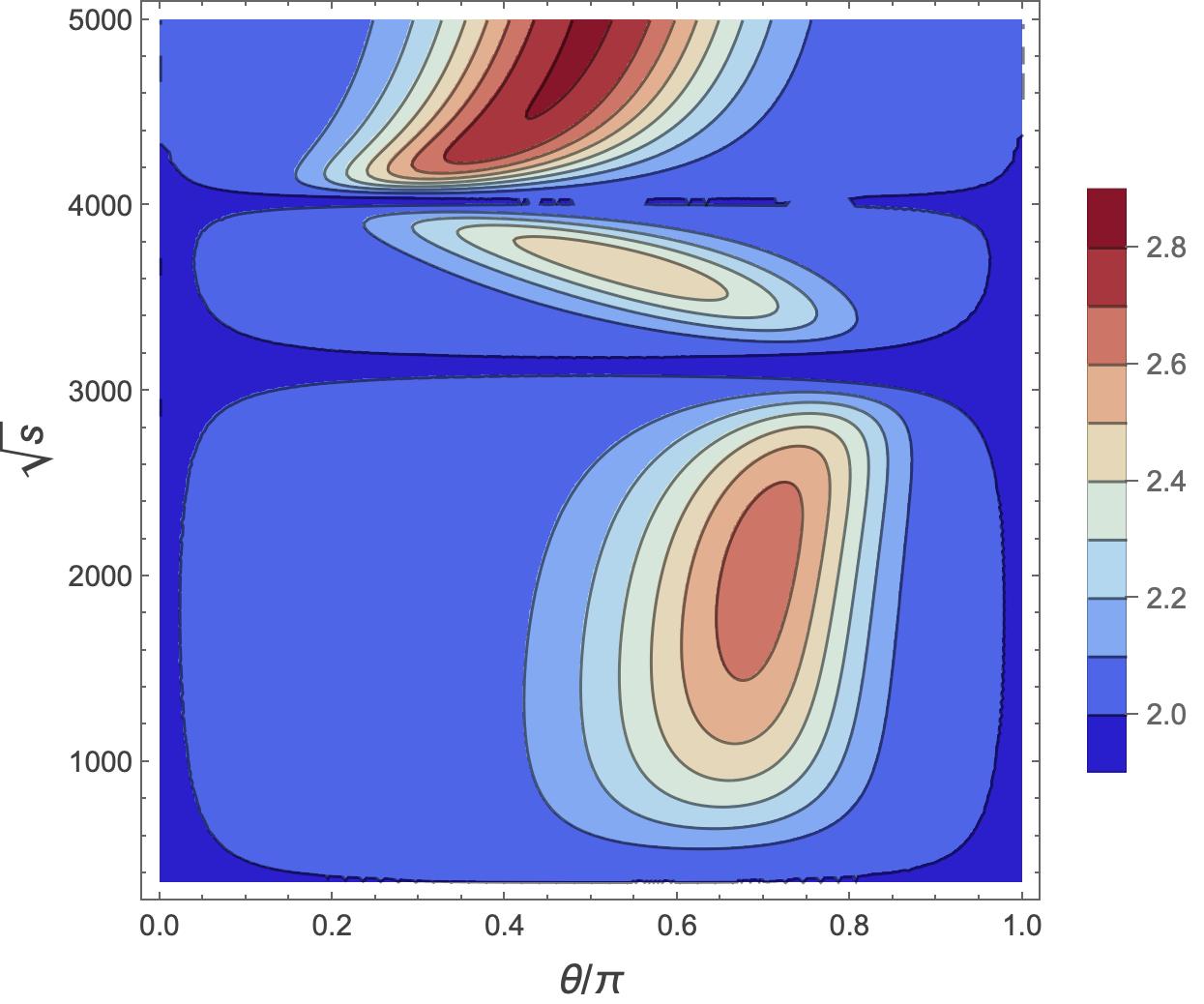}
  \end{minipage}
  \begin{minipage}[b]{0.45\linewidth}
    \centering
    \includegraphics[keepaspectratio, scale=0.3]{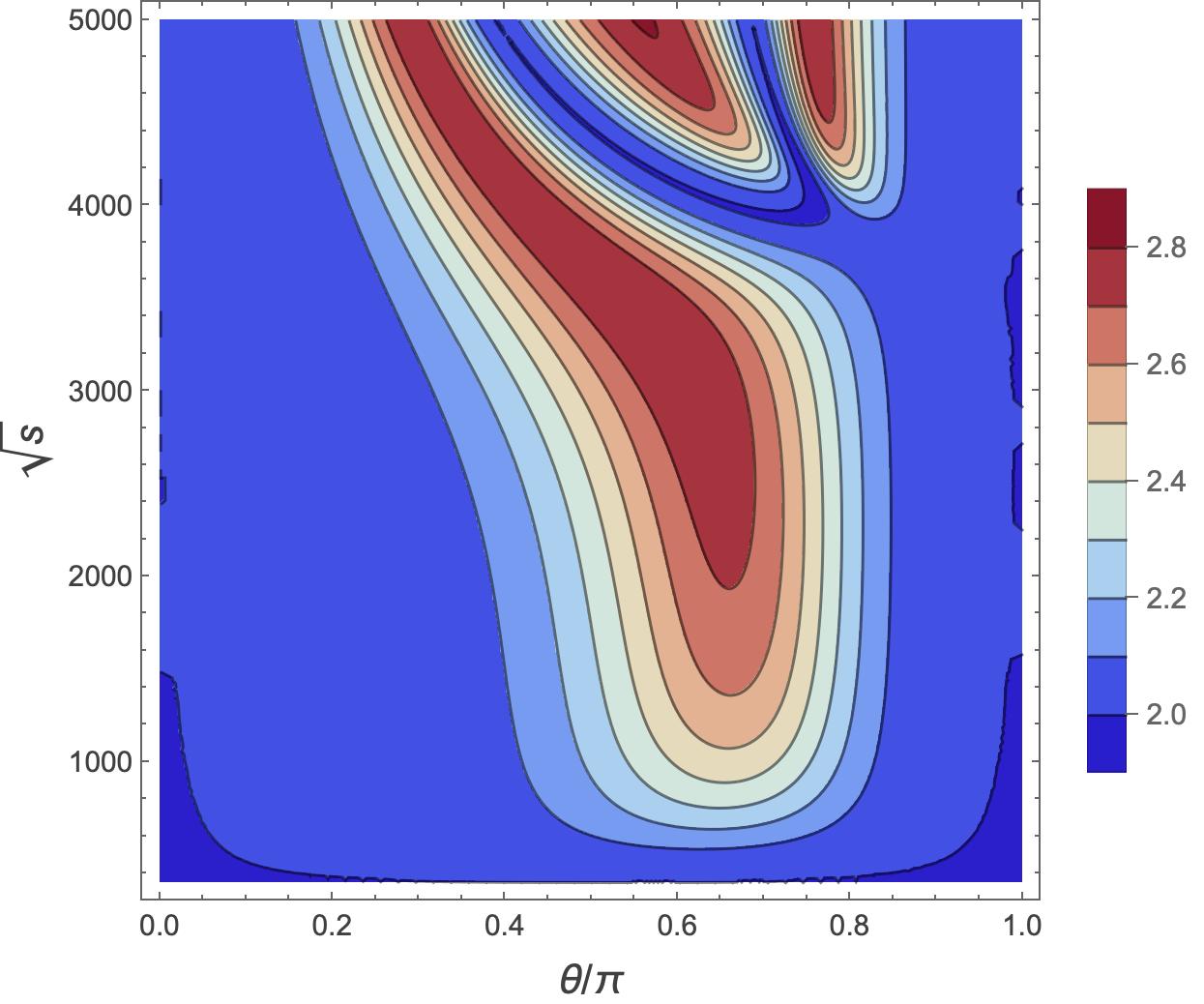}
  \end{minipage}
\caption{Contours of the CHSH Bell parameter in the $(\sqrt{s},\theta/\pi)$ plane.
The panel assignments are the same as in Fig.~\ref{fig:EmarkerContours_sqrts}
 for the entanglement marker.}
\label{fig:BContours_sqrts}
\end{center}
\end{figure}

\begin{figure}[htbp]
\begin{center}
  \begin{minipage}[b]{0.45\linewidth}
    \centering
    \includegraphics[keepaspectratio, scale=0.3]{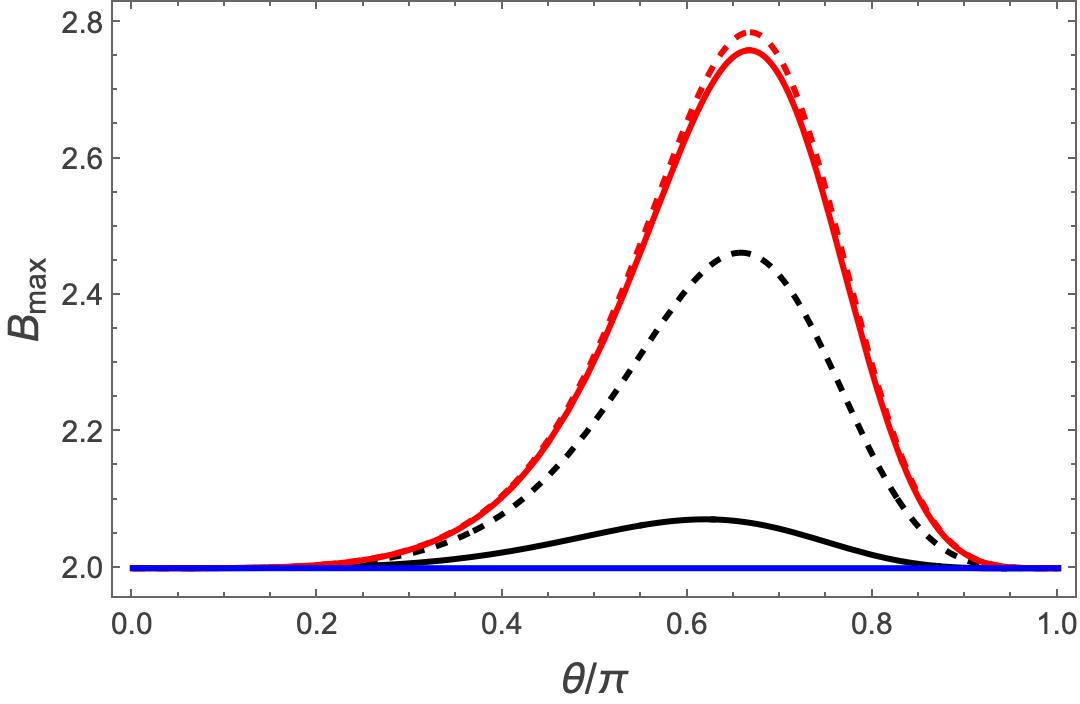}
  \end{minipage}
  \begin{minipage}[b]{0.45\linewidth}
    \centering
    \includegraphics[keepaspectratio, scale=0.3]{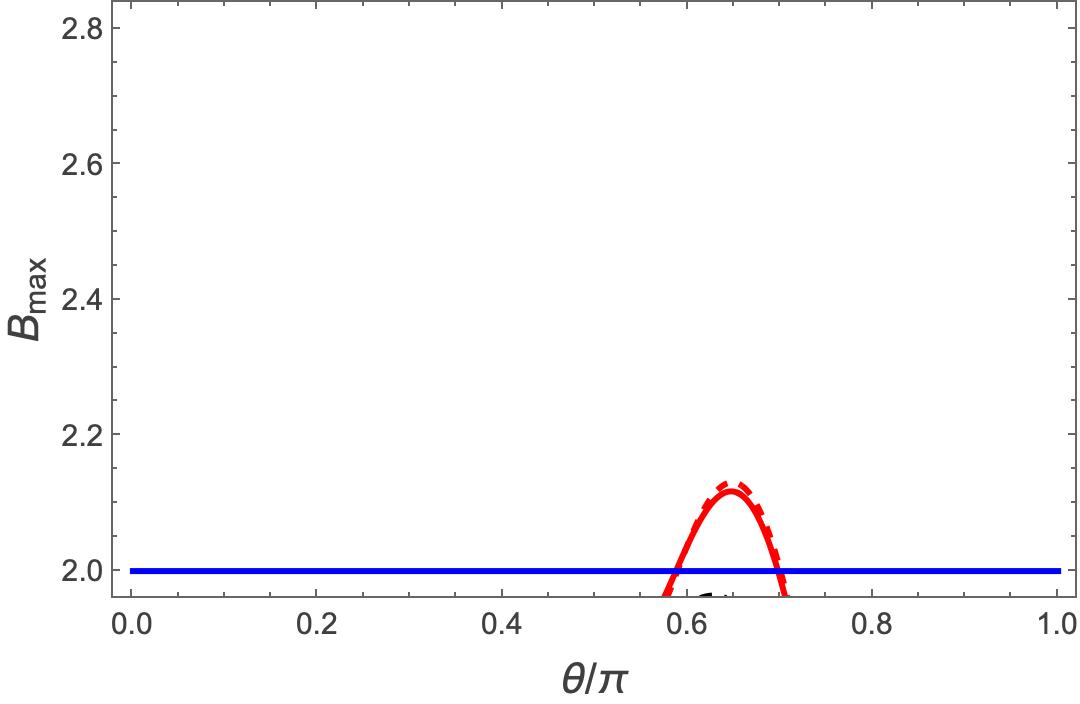}
  \end{minipage} \par\medskip
  \begin{minipage}[b]{0.45\linewidth}
    \centering
    \includegraphics[keepaspectratio, scale=0.3]{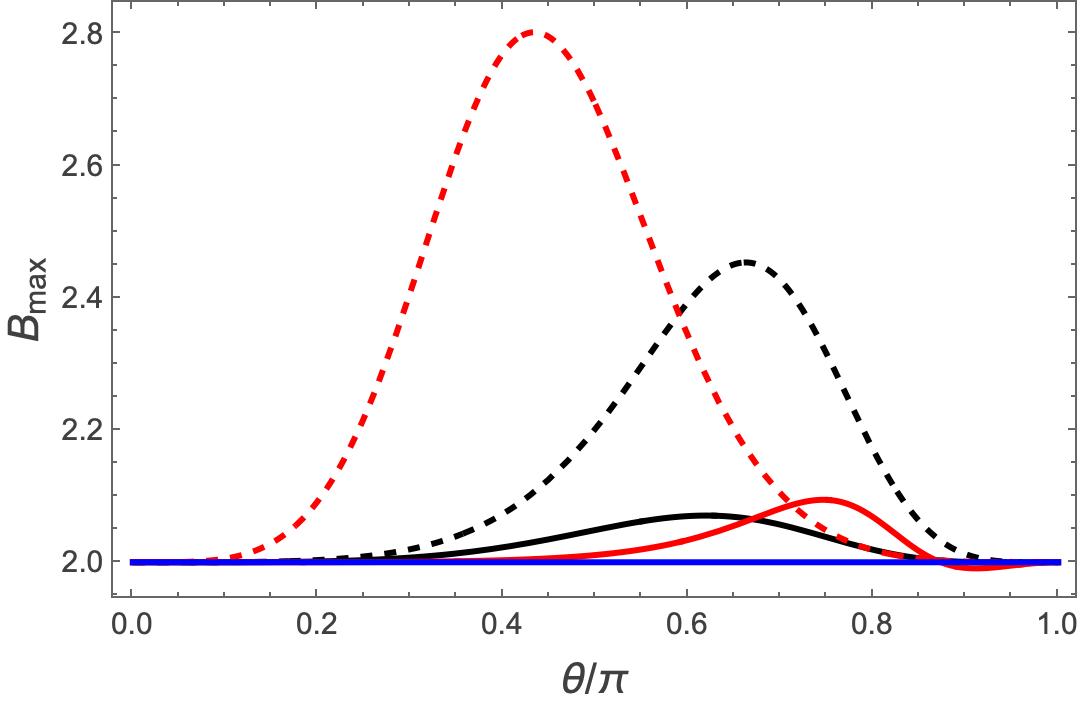}
  \end{minipage}
  \begin{minipage}[b]{0.45\linewidth}
    \centering
    \includegraphics[keepaspectratio, scale=0.3]{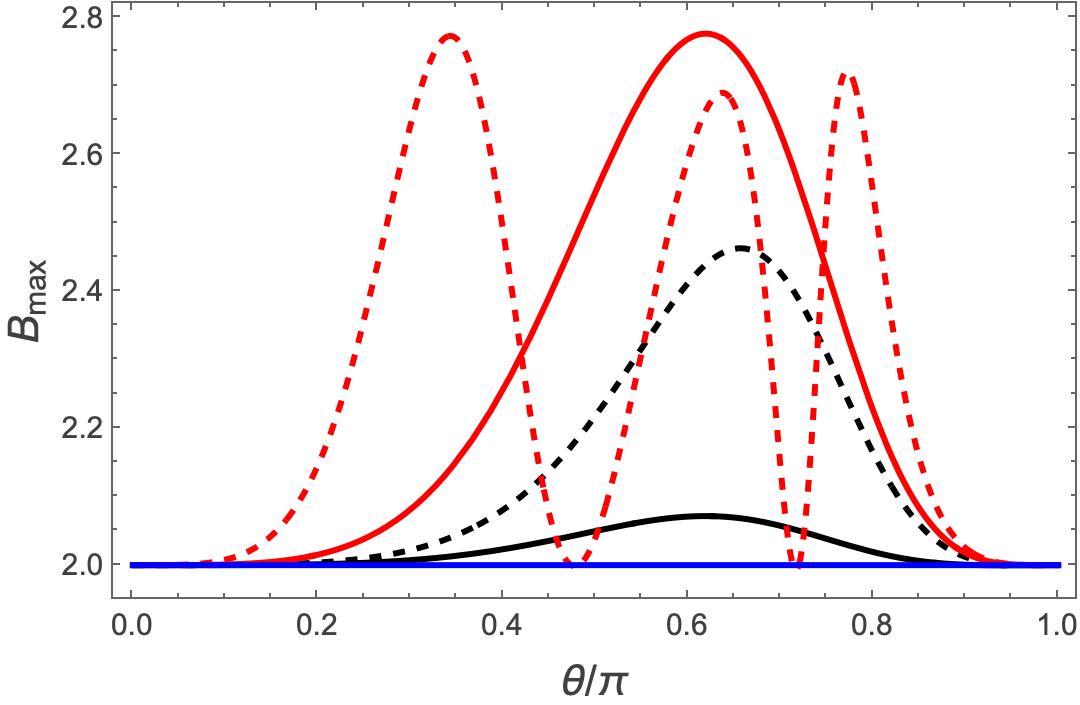}
  \end{minipage}
\caption{Angular dependence of the CHSH Bell parameter at fixed center-of-mass energies.
The panel assignments are the same as in Fig.~\ref{fig:EmarkerContoursSlices}.
The solid black, dashed black, solid red, and dashed red curves correspond to
$\sqrt{s}=500~\mathrm{GeV}$, $1000~\mathrm{GeV}$, $3000~\mathrm{GeV}$, and $4500~\mathrm{GeV}$, respectively.
The horizontal blue line marks the classical CHSH bound $B_{\max}=2$.
}
\label{fig:BContoursSlices}
\end{center}
\end{figure}

We finally turn to the maximal CHSH Bell parameter,
 which probes a more restrictive aspect of quantum correlations than
 either the entanglement marker or the concurrence.
It is therefore interesting to examine whether the qualitative
 differences among the scalar, vector, and RS scenarios persist
 at the level of Bell nonlocality.
The plots of the CHSH Bell parameter are
 shown in Fig.~\ref{fig:BContours_sqrts} and Fig.~\ref{fig:BContoursSlices}.
The contour plots in Fig.~\ref{fig:BContours_sqrts} exhibit qualitative features similar to those 
 of the entanglement marker and the concurrence,
 indicating that the qualitative differences among the scalar,
 vector, and RS scenarios persist across all three observables.

A similar correspondence is observed in the energy slices in Fig.~\ref{fig:BContoursSlices}.
Despite probing a more restrictive aspect of quantum correlations, the Bell parameter 
 retains the same qualitative distinctions among the scalar, vector, and RS scenarios that 
 were observed in the entanglement marker and concurrence.
In the SM case shown in the top-left panel of Fig.~\ref{fig:BContoursSlices},
 the Bell parameter increases with energy and develops a single broad peak in the angular
 distribution, leading to Bell-inequality violation at sufficiently high energies.
The scalar scenario in the top-right panel of Fig.~\ref{fig:BContoursSlices}
 shows a much more limited pattern of Bell-inequality violation.
For $\sqrt{s}=500$ and $1000~\mathrm{GeV}$, the Bell parameter stays below 
 the classical bound over the entire angular range.
Only at higher energies, $\sqrt{s}=3000$ and $4500~\mathrm{GeV}$, does 
 $B_{\max}$ exceed 2, and then only within a narrow interval of intermediate scattering angles.
The $U(1)_{B-L}$ model shown in the bottom-left panel of
Fig.~\ref{fig:BContoursSlices}
exhibits a single peak whose angular position evolves with
$\sqrt{s}$.
The peak is strongly suppressed at
$\sqrt{s}=3~{\rm TeV}$ and increases at higher energies,
following the behavior already observed in the concurrence.
In contrast, the RS scenario shown in the bottom-right panel of Fig.~\ref{fig:BContoursSlices}
 exhibits a broad enhancement at lower energies, while at higher energies the angular profile becomes 
 increasingly structured.
In particular, for $\sqrt{s}=4500~\mathrm{GeV}$, multiple local maxima
 emerge, reflecting the characteristic angular structure of the spin-2
 KK-graviton exchange.
These results further demonstrate that the characteristic differences among the scalar, 
vector, and RS scenarios are consistently reflected in all three quantum-information observables.

Taken together, the results for the entanglement marker, concurrence,
 and Bell parameter reveal a common pattern.
The scalar scenario primarily suppresses the SM quantum correlations,
while the $U(1)_{B-L}$ scenario is characterized by the
energy-dependent evolution of the angular extrema governed by the
effective chiral combinations, together with the characteristic
weakening of the entanglement caused by the cancellation condition.
The RS scenario instead develops characteristic angular structures
associated with spin-2 KK-graviton exchange, leading to qualitatively
distinct signatures absent in the scalar and vector benchmarks considered
here.
The fact that the same qualitative distinctions among the scalar, vector, 
 and RS scenarios are reflected in all three observables suggests that
 these distinctions are closely linked to the properties of the underlying 
 new particles and their interaction patterns. 

We note that all the numerical results shown in this section were confirmed by automated
 simulation of event-by-event spin-density matrices~\cite{Durupt:2025wuk} 
 in the \texttt{MadGraph5\_aMC@NLO} (\texttt{MadGraph} hereafter) framework~\cite{Alwall:2014hca}.
As a reference, we show Tables~\ref{tab:spin_corr_matrix} and~\ref{tab:Dcoeff_comparison}, 
where we compare
the diagonal components of the spin-correlation matrix $C_{ij}$ and the entanglement
markers $D^{(i)}$ obtained from the analytical calculation (Theory) and from the numerical simulation by \texttt{MadGraph}.
Here, for simplicity, we isolate the contributions of each new-physics scenario from the SM contribution,
and show the results for the benchmarks defined in Table~\ref{Table:benchmark} at $\sqrt{s}=3000$~GeV 
with $10^5$ generated events for simulation.
For simulations in the scalar-mediator (spin-0) model and the RS (spin-2) model, we employ the \texttt{Higgs Charactersisation} (\texttt{HC}) model~\cite{Artoisenet:2013puc}, while
for simulations in the $U(1)_{B-L}$ (spin-1) model, we use the \texttt{U1XGeneric} model~\cite{Das:2021esm}.
For the RS case, we only include the contribution from the first KK graviton mode for the comparison tables.
We find good agreement between the two approaches within statistical uncertainties,
providing a nontrivial cross-check of the analytical framework.

 \begin{table}
\centering
\resizebox{\linewidth}{!}{
\begin{tabular}{c|rr|rr|rr}
\hline
Model
& \multicolumn{2}{c|}{$C_{rr}$}
& \multicolumn{2}{c|}{$C_{nn}$}
& \multicolumn{2}{c}{$C_{kk}$} \\
& Theory & \texttt{MadGraph} & Theory & \texttt{MadGraph} & Theory & \texttt{MadGraph} \\
\hline
SM
& $0.300$ & $0.298$
& $-0.290$ & $-0.287$
& $0.989$ & $0.990$ \\

scalar-mediator
& $-0.007$ & $-0.007$
& $-0.007$ & $-0.007$
& $-1.000$ & $-1.000$ \\

$U(1)_{B-L}$
& $0.503$ & $0.502$
& $-0.490$ & $-0.489$
& $0.987$ & $0.987$ \\

RS
& $-0.156$ & $-0.158$
& $0.174$ & $0.176$
& $0.982$ & $0.982$ \\
\hline
\end{tabular}
}
\caption{Comparison of the diagonal components of the spin-correlation matrix
between the analytical calculation (Theory) and the numerical simulation by \texttt{MadGraph}.}
\label{tab:spin_corr_matrix}
\end{table}
 
\begin{table}
\centering
\resizebox{\linewidth}{!}{
\begin{tabular}{c|rr|rr|rr|rr}
\hline
Model
& \multicolumn{2}{c|}{$D^{(1)}$}
& \multicolumn{2}{c|}{$D^{(k)}$}
& \multicolumn{2}{c|}{$D^{(r)}$}
& \multicolumn{2}{c}{$D^{(n)}$} \\
& Theory & \texttt{MadGraph} & Theory & \texttt{MadGraph} & Theory & \texttt{MadGraph} & Theory & \texttt{MadGraph} \\
\hline
SM
& $0.333$ & $0.333$
& $0.326$ & $0.326$
& $-0.133$ & $-0.135$
& $-0.526$ & $-0.525$\\

scalar-mediator
& $-0.338$ & $-0.338$
& $-0.329$ & $-0.329$
& $0.333$ & $0.333$
& $0.333$ & $0.333$ \\

$U(1)_{B-L}$
& $0.333$ & $0.333$
& $0.325$ & $0.325$
& $0.002$ & $0.002$
& $-0.660$ & $-0.660$ \\

RS
& $0.333$ & $0.333$
& $0.322$ & $0.322$
& $-0.438$ & $-0.439$
& $-0.217$ & $-0.216$ \\
\hline
\end{tabular}
}
\caption{Comparison of the entanglement markers $D^{(i)}$ obtained from the analytical calculation (Theory) and the numerical simulation by \texttt{MadGraph}.}
\label{tab:Dcoeff_comparison}
\end{table}

\section{Discussion and conclusions}
\label{sec:conclusion}
In this work, we have performed a systematic study of quantum entanglement 
and Bell-inequality violation in $l^-l^+ \to t\bar t$ $(l=e,\mu)$ at future lepton colliders
 such as the ILC and multi-TeV muon colliders, 
within the SM and three representative extensions involving scalar, vector,
 and tensor-mediated interactions.
Working in a unified density-matrix framework, we evaluated the entanglement marker, 
the concurrence, and the maximal CHSH Bell parameter on equal footing.

The central result of this work is that the scalar, vector, and RS scenarios
 leave qualitatively distinct fingerprints in quantum-information observables.
Rather than merely enhancing or suppressing entanglement, the new interactions produce 
 qualitatively distinct patterns in the quantum-information observables, reflecting the 
 properties of the underlying new particles and their interaction patterns.
The scalar mediator leads to an overall reduction of entanglement,
 reflecting its tendency to enhance like-helicity top-pair configurations.
The absence of strong interference with the SM amplitudes further ensures
 that the overall topology remains similar to the SM pattern.
In contrast, the $U(1)_{B-L}$ model is characterized by the
energy-dependent evolution of the angular extrema together with a
characteristic weakening of the entanglement, leading to a
characteristic reshaping of the entanglement pattern in the
$(\theta,\sqrt{s})$ plane.
The RS scenario exhibits the most distinctive behavior,
 developing characteristic angular structures associated with spin-2
 KK-graviton exchange that are absent in the scalar and vector
 benchmarks considered here.

We further find that the CHSH parameter provides information complementary to that of the entanglement marker and concurrence. 
In the SM, as well as in the $U(1)_{B-L}$ and RS scenarios, the maximal CHSH parameter exceeds the classical bound 
 $B_{\max}=2$ over broad angular regions, indicating persistent Bell-inequality violation throughout much of the accessible phase space. 
By contrast, in the scalar-mediator model the violation is much more limited:
for $\sqrt{s}=500$ and $1000~\mathrm{GeV}$, the CHSH parameter stays below the classical bound over the entire angular range,
whereas clear violation appears only at higher center-of-mass energies, where it is confined to narrow angular intervals.
This behavior reflects the different ways in which the interactions considered here 
 reshape the underlying quantum-correlation structure.
The vector and RS scenarios exhibit Bell-inequality violation over broader regions
 of phase space, whereas the scalar contribution produces a more limited pattern
 that becomes visible mainly at higher energies.
Taken together, the entanglement marker, concurrence, and CHSH parameter therefore 
 provide complementary diagnostics of the properties of new particles and their interaction patterns.

Future lepton colliders provide a uniquely clean environment for probing 
 quantum entanglement with $t\bar t$ final state. 
In particular, polarized beams would allow direct control over the initial-state 
helicity configuration and could significantly enhance the sensitivity to the 
 chiral and Lorentz structure of new interactions \cite{Fang:2026ddi}. 
Incorporating realistic reconstruction and detector effects will be essential 
 for assessing the experimental feasibility of measuring these quantum-information 
 observables in practice. 
Such studies will help clarify the role of quantum-information observables as 
probes of new physics at future lepton colliders.

\vspace{10mm}

\noindent{\Large \bf Appendix}

\appendix

\section{Chiral gauge coupling dependence of the entanglement observables}
\label{app:chiral_vector}

To clarify the origin of the energy-dependent shift of the
 entanglement extremum observed in the $U(1)_{B-L}$ scenario as discussed in Sec. \ref{sec:numerics},
 we consider the production of a massless fermion pair through a
 generic neutral vector boson with independent left- and
 right-chiral couplings.
The interaction Lagrangian is written as
\begin{align}
 {\cal L}_{\rm int}
 =
 V_\mu
 \left[
 \bar e\gamma^\mu
 \left(g_L^e P_L+g_R^e P_R\right)e
 +
 \bar t\gamma^\mu
 \left(g_L^t P_L+g_R^t P_R\right)t
 \right],
 \label{eq:generic_vector_lagrangian}
\end{align}
where
\begin{align}
 P_L=\frac{1-\gamma_5}{2},
 \qquad
 P_R=\frac{1+\gamma_5}{2}.
\end{align}
The couplings are taken to be real. We neglect the electron and top
masses in order to isolate the dependence of the angular distribution
on the chiral structure of the interaction. The overall propagator
and normalization factors cancel in the normalized density matrix and
will therefore be omitted below.

The normalized density matrix is
\begin{align}
 \rho(\theta)
 =
 \frac{1}{N(\theta)}
 \begin{pmatrix}
 A(\theta)&0&0&B(\theta)\\
 0&0&0&0\\
 0&0&0&0\\
 B(\theta)&0&0&D(\theta)
 \end{pmatrix},
 \label{eq:generic_vector_normalized_density}
\end{align}
where
\begin{align}
 N(\theta)&=A(\theta)+D(\theta),\\
 A(\theta)
 &=
 4\left(g_R^t\right)^2
 \left[
 \left(g_R^e\right)^2
 \cos^4\frac{\theta}{2}
 +
 \left(g_L^e\right)^2
 \sin^4\frac{\theta}{2}
 \right],
 \label{eq:generic_vector_A}
 \\
 D(\theta)
 &=
 4\left(g_L^t\right)^2
 \left[
 \left(g_L^e\right)^2
 \cos^4\frac{\theta}{2}
 +
 \left(g_R^e\right)^2
 \sin^4\frac{\theta}{2}
 \right],
 \label{eq:generic_vector_D}
 \\
 B(\theta)
 &=
 \left[
 \left(g_L^e\right)^2+\left(g_R^e\right)^2
 \right]
 g_L^t g_R^t\sin^2\theta.
 \label{eq:generic_vector_B}
\end{align}

From the density matrix (\ref{eq:generic_vector_normalized_density}), we have the entanglement marker as
\begin{align}
 D_{\min}(\theta)
 &=
 -\frac{1}{3}
 \left[
 1+\frac{4|B(\theta)|}{N(\theta)}
 \right].
 \label{eq:generic_vector_Dmin}
\end{align}
Thus, the entanglement condition (\ref{Dentanglement})
 is satisfied whenever the off-diagonal element
 $B(\theta)$ is nonzero.

The concurrence can also be evaluated analytically. 
For the density matrix in Eq.~\eqref{eq:generic_vector_normalized_density},
 we have
\begin{align}
 C(\theta)
 =
 \frac{2|B(\theta)|}{N(\theta)}.
 \label{eq:generic_vector_concurrence}
\end{align}
Consequently, the entanglement marker and concurrence satisfy the
simple relation
\begin{align}
 D_{\min}(\theta)
 =
 -\frac{1+2C(\theta)}{3}.
 \label{eq:Dmin_concurrence_relation}
\end{align}
The two observables therefore contain the same angular information
in the present massless single-vector-exchange limit. In particular,
the minimum of $D_{\min}$ occurs at the same angle as the maximum of
the concurrence.

The dependence of this angle on the chiral couplings can be obtained
explicitly. Introducing
\begin{align}
 P
 &=
 \left(g_R^t g_R^e\right)^2
 +
 \left(g_L^t g_L^e\right)^2,
 \label{eq:generic_vector_P}
 \\
 Q
 &=
 \left(g_R^t g_L^e\right)^2
 +
 \left(g_L^t g_R^e\right)^2,
 \label{eq:generic_vector_Q}
\end{align}
and defining
\begin{align}
 x=\cos\theta,
\end{align}
the normalization factor becomes
\begin{align}
 N(x)
 =
 P(1+x)^2+Q(1-x)^2,
 \label{eq:generic_vector_N_PQ}
\end{align}
whereas
\begin{align}
 |B(x)|
 =
 R(1-x^2),
 \qquad
 R=
 \left[
 \left(g_L^e\right)^2+\left(g_R^e\right)^2
 \right]
 \left|g_L^t g_R^t\right|.
 \label{eq:generic_vector_R}
\end{align}
It follows that
\begin{align}
 C(x)
 =
 \frac{2R(1-x^2)}
 {P(1+x)^2+Q(1-x)^2}.
 \label{eq:generic_vector_concurrence_PQ}
\end{align}
The nontrivial extremum of (\ref{eq:generic_vector_concurrence_PQ}) in the physical angular region is located at
\begin{align}
 \cos\theta_{\rm ext}
 =
 \frac{\sqrt{Q}-\sqrt{P}}
 {\sqrt{P}+\sqrt{Q}}
.
 \label{eq:generic_vector_extremum}
\end{align}
At this angle, the concurrence reaches
\begin{align}
 C_{\rm max}
 =
 \frac{R}{\sqrt{PQ}},
 \label{eq:generic_vector_Cmax}
\end{align}
and the entanglement marker takes the minimum value
\begin{align}
 D_{\min}^{\rm(min)}
 =
 -\frac{1}{3}
 \left(
 1+\frac{2R}{\sqrt{PQ}}
 \right).
 \label{eq:generic_vector_Dmin_minimum}
\end{align}

Equation~\eqref{eq:generic_vector_extremum} shows explicitly that the
angular position of the entanglement structure is controlled by the
relative strengths of the chiral couplings. The distribution is
symmetric about $\theta=\pi/2$ when
\begin{align}
 P=Q.
 \label{eq:generic_vector_symmetric_condition}
\end{align}
Indeed,
\begin{align}
 P-Q
 =
 \left[
 \left(g_R^t\right)^2-\left(g_L^t\right)^2
 \right]
 \left[
 \left(g_R^e\right)^2-\left(g_L^e\right)^2
 \right].
 \label{eq:generic_vector_PminusQ}
\end{align}
Therefore, $P=Q$ if the coupling to either the initial or final
fermion is vector-like in magnitude. In particular, for the
$U(1)_{B-L}$ interaction,
\begin{align}
 g_L^e=g_R^e,
 \qquad
 g_L^t=g_R^t,
 \label{eq:BL_vectorlike}
\end{align}
and one obtains
\begin{align}
 \theta_{\rm ext}=\frac{\pi}{2}.
\end{align}
By contrast, if $P>Q$, the extremum is shifted to
$\theta>\pi/2$, while for $P<Q$ it is shifted to
$\theta<\pi/2$.

\begin{figure}[t]
 \centering
 \includegraphics[width=0.55\linewidth]{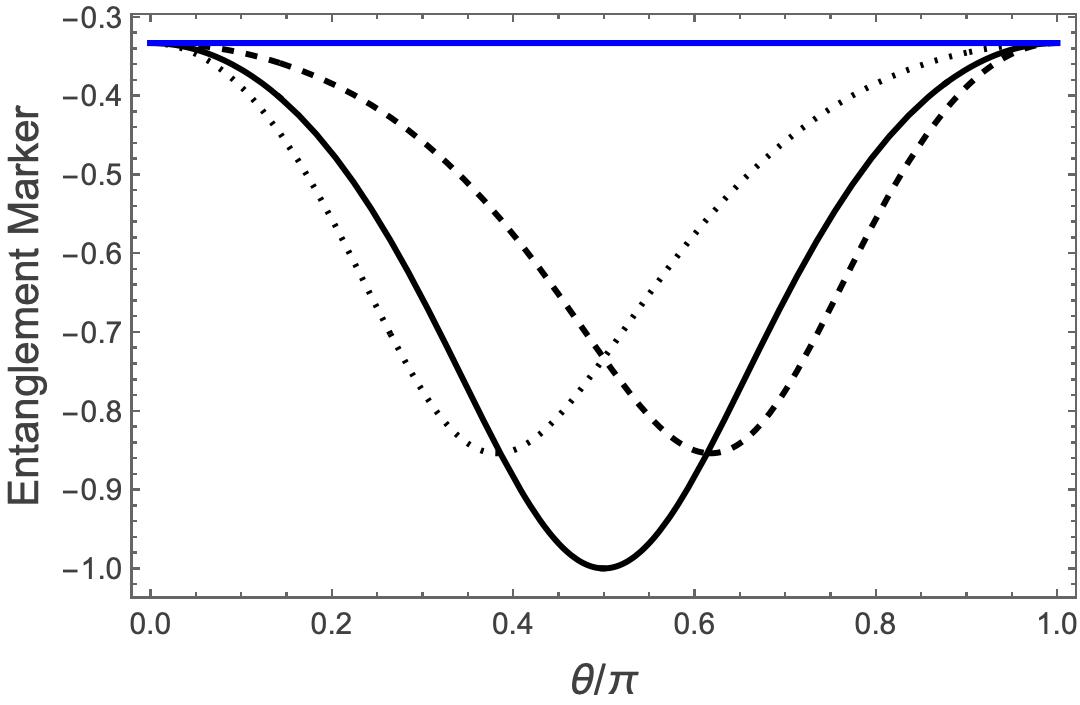}
 \caption{
Angular dependence of the entanglement marker for several
 representative choices of chiral couplings.
The blue horizontal line denotes the separability bound, $D_{\rm min}=-1/3$.
The solid curve corresponds to vector-like couplings $P=Q,~(g^e_R=g^e_L=g^t_R=g^t_L=1$),
 while the dashed and dotted curves illustrate the cases
$P>Q~(g^l_R=g^t_R=3, g^l_L=g^t_L=1)$ and $P<Q~(g^l_R=g^t_R=1, g^l_L=g^t_L=3)$, respectively.
The minimum shifts according to
Eq.~\eqref{eq:generic_vector_extremum},
 demonstrating that its angular position is determined by the relative
 strengths of the left- and right-chiral couplings.
 couplings.
 } 
 \label{fig:generic_vector_shift}
\end{figure}

We show the dependence of the entanglement marker on the chiral couplings
 in Fig.~\ref{fig:generic_vector_shift}.
For vector-like couplings,
 $P=Q$, and the minimum is located at $\theta=\pi/2$.
When $P>Q$, the minimum is shifted towards
 $\theta>\pi/2$, whereas
$P<Q$ shifts the minimum towards
 $\theta<\pi/2$, in complete agreement with
Eq.~\eqref{eq:generic_vector_extremum}.

These results provide a simple interpretation of the
 energy-dependent angular structures found in the full
 $U(1)_{B-L}$ calculation. 
At relatively low energies, the
 distribution is dominated by the SM amplitudes and reflects their
 chiral asymmetry. 
In the energy region where the $Z'$ contribution
 dominates, the vector-like $U(1)_{B-L}$ couplings drive the
 entanglement extremum toward the symmetric position
 $\theta/\pi=1/2$. 
The energy-dependent movement of the entanglement extrema and
 the weakened-entanglement regions observed in the intermediate-energy region 
 require the coexistence of chiral and vector-like interactions. 
These features are analyzed in Appendix~\ref{app:vector_interference}.

\section{Analytic interpretation of the intermediate-energy entanglement pattern in the $U(1)_{B-L}$ model}
\label{app:vector_interference}

As discussed in Sec.~\ref{sec:numerics},
the $U(1)_{B-L}$ model exhibits two characteristic features in the
intermediate-energy region.
The entanglement extrema move with the collision energy, while
the entanglement is significantly weakened around
$\sqrt{s}\simeq3~{\rm TeV}$ and
$4~{\rm TeV}$, leading to the weakened-entanglement regions in
Fig.~\ref{fig:EmarkerContours_sqrts}.
To clarify the origin of this behavior, we consider a simplified model containing a
massless neutral vector boson $A_\mu$ with generic chiral couplings and
a massive neutral vector boson $X_\mu$ with vector-like couplings.
The former mimics the chiral structure of the SM $Z$ boson, with its
mass neglected, whereas the latter mimics the vector-like $Z'$ boson
of the $U(1)_{B-L}$ model.  
The interaction Lagrangian is
\begin{align}
{\cal L}_{\rm int}
={}&
A_\mu
\left[
\bar e\gamma^\mu
\left(g_L^lP_L+g_R^lP_R\right)e
+
\bar t\gamma^\mu
\left(g_L^tP_L+g_R^tP_R\right)t
\right]
\nonumber\\
&+
g_{B-L}X_\mu
\left(
Q_{B-L}^l\,\bar e\gamma^\mu e
+
Q_{B-L}^t\,\bar t\gamma^\mu t
\right).
\label{eq:interference_lagrangian}
\end{align}
We neglect the lepton and top-quark masses, as in
Appendix~\ref{app:chiral_vector}, and also neglect the width of the
massive vector boson.

The normalized density matrix takes the form
\begin{align}
 \rho(s,\theta)
 =
 \frac{1}{N(s,\theta)}
 \begin{pmatrix}
 A(s,\theta)&0&0&B(s,\theta)\\
 0&0&0&0\\
 0&0&0&0\\
 B(s,\theta)&0&0&D(s,\theta)
 \end{pmatrix},
 \label{eq:interference_density_matrix}
\end{align}
where
\begin{align}
 A(s,\theta)
 &=
 4\left[
 G_{RR}^2(s)\cos^4\frac{\theta}{2}
 +
 G_{LR}^2(s)\sin^4\frac{\theta}{2}
 \right],
 \\
 D(s,\theta)
 &=
 4\left[
 G_{LL}^2(s)\cos^4\frac{\theta}{2}
 +
 G_{RL}^2(s)\sin^4\frac{\theta}{2}
 \right],
 \\
 B(s,\theta)
 &=
 \left[
 G_{RR}(s)G_{RL}(s)
 +
 G_{LR}(s)G_{LL}(s)
 \right]\sin^2\theta,
 \\
 N(s,\theta)
 &=
 A(s,\theta)+D(s,\theta),
 \label{eq:interference_ABD}
\end{align}
with
\begin{align}
 G_{RR}(s)
 &=
 g_R^lg_R^t-\chi(s),
 &
 G_{LR}(s)
 &=
 g_L^lg_R^t-\chi(s),  \label{eq:G_chiral_channels1}
 \\
 G_{LL}(s)
 &=
 g_L^lg_L^t-\chi(s),
 &
 G_{RL}(s)
 &=
 g_R^lg_L^t-\chi(s),
 \label{eq:G_chiral_channels2}
\end{align}
and
\begin{align}
 \chi(s)
 =
 \frac{g_{B-L}^2}{3}
 \frac{s}{s-M_X^2},
 \label{eq:chi_definition}
\end{align}
where $M_X$ is a mass of the $X$-boson.
Here the factor $1/3$ follows from
$Q_{B-L}^lQ_{B-L}^t=-1/3$, with an overall common sign absorbed into
the definition of the four channel coefficients.
The quantities $G_{RR}(s)$, $G_{LR}(s)$, $G_{LL}(s)$, and
$G_{RL}(s)$ may be regarded as effective chiral couplings, since
they incorporate both the original chiral couplings and the
energy-dependent vector-like contribution $\chi(s)$.

For this density matrix, the concurrence and entanglement marker are, respectively, 
\begin{align}
 C(s,\theta)
 &=
 \frac{2|B(s,\theta)|}{N(s,\theta)},
 \\
 D_{\min}(s,\theta)
 &=
 -\frac{1}{3}
 \left[
 1+\frac{4|B(s,\theta)|}{N(s,\theta)}
 \right]
 =
 -\frac{1+2C(s,\theta)}{3}.
 \label{eq:interference_entanglement}
\end{align}
The weakening of the entanglement is therefore controlled by the
ratio $|B|/N$.  For $\sin\theta\neq0$, the concurrence vanishes when
\begin{align}
 G_{RR}(s)G_{RL}(s)
 +
 G_{LR}(s)G_{LL}(s)
 =0.
 \label{eq:interference_zero_condition}
\end{align}
This can occur for suitable combination of gauge couplings and collision energies.
In such a case, the concurrence vanishes and the entanglement marker reaches the separability bound.

Expanding Eq.~\eqref{eq:interference_zero_condition} in powers of
$\chi$, we obtain
\begin{align}
 2\chi^2-S\chi+T=0,
 \label{eq:chi_quadratic}
\end{align}
where
\begin{align}
 S
 &=
 \left(g_L^l+g_R^l\right)
 \left(g_L^t+g_R^t\right),
 \\
 T
 &=
 \left[
 \left(g_L^l\right)^2+\left(g_R^l\right)^2
 \right]g_L^tg_R^t.
 \label{eq:ST_definitions}
\end{align}
The solutions are
\begin{align}
 \chi_\pm
 =
 \frac{
 S\pm\sqrt{S^2-8T}
 }{4}.
 \label{eq:chi_roots}
\end{align}
A real suppression point exists only if
\begin{align}
 S^2-8T\geq0.
 \label{eq:real_suppression_condition}
\end{align}
Using Eq.~\eqref{eq:chi_definition}, the corresponding collision
energies are
\begin{align}
 s_\pm
 &=
 \frac{\chi_\pm}
 {\chi_\pm-g_{BL}^2/3}M_X^2,
 \\
 \sqrt{s_\pm}
 &=
 M_X
 \sqrt{
 \frac{\chi_\pm}
 {\chi_\pm-g_{BL}^2/3}
 }.
 \label{eq:suppression_energy}
\end{align}
Conversely, for a target energy $\sqrt{s_0}$,
\begin{align}
 M_X
 =
 \sqrt{s_0}
 \sqrt{
 1-\frac{g_{B-L}^2}{3\chi_\pm}
 }.
 \label{eq:mediator_mass_from_target}
\end{align}
A physical solution requires
$\chi_\pm/(\chi_\pm-g_{BL}^2/3)>0$.

If the massless interaction is also vector-like,
\begin{align}
 g_L^l=g_R^l,
 \qquad
 g_L^t=g_R^t,
\end{align}
the four coefficients in Eqs.~\eqref{eq:G_chiral_channels1} and \eqref{eq:G_chiral_channels2} become
identical.  Away from a zero of the total production amplitude, their
common energy-dependent factor cancels from the normalized density
matrix.  
Consequently, all elements of the density matrix acquire the same
energy-dependent factor, which cancels in the normalized density
matrix. Therefore, purely vector-like interactions do not
generate an energy-dependent suppression of the normalized
entanglement observables.

As a concrete example, let us take
\begin{align}
 M_X=5~{\rm TeV},
 \qquad
 g_{B-L}=1,
 \qquad
 g_L^l=g_R^l=1,
 \qquad
 g_L^t=-\frac{16}{27},
 \qquad
 g_R^t=-\frac{3}{16}.
 \label{eq:example_chiral_couplings}
\end{align}
For this choice,
\begin{align}
 \chi(3~{\rm TeV})
 &=
 -\frac{3}{16},
 \qquad
 \chi(4~{\rm TeV})
 =
 -\frac{16}{27}.
\end{align}
The off-diagonal element becomes
\begin{align}
 B(s,\theta)
 =
 2
 \left(
 -\frac{3}{16}-\chi(s)
 \right)
 \left(
 -\frac{16}{27}-\chi(s)
 \right)
 \sin^2\theta.
\end{align}
The first factor vanishes at $\sqrt{s}=3~{\rm TeV}$, while the second
factor vanishes at $\sqrt{s}=4~{\rm TeV}$.  Consequently,
\begin{align}
 C(s,\theta)=0,
 \qquad
 D_{\min}(s,\theta)=-\frac{1}{3}
\end{align}
at these two energies for $\sin\theta\neq0$, whereas the diagonal
normalization remains nonzero.  This example explicitly demonstrates
that the weakening of the entanglement can arise from cancellations
among the chiral-channel contributions to the off-diagonal
density-matrix element.

The above analysis also provides a simple interpretation of the
motion of the entanglement extrema.
In particular, it explains the characteristic behavior observed
in the intermediate-energy region, where the SM-like and vector-like
contributions compete.
Introducing
\begin{align}
 P(s)
 &=
 G_{RR}^2(s)+G_{LL}^2(s),  \label{eq:PQR_energy_dependent1}
 \\
 Q(s)
 &=
 G_{LR}^2(s)+G_{RL}^2(s),  \label{eq:PQR_energy_dependent2}
 \\
 R(s)
 &=
 \left|
 G_{RR}(s)G_{RL}(s)
 +
 G_{LR}(s)G_{LL}(s)
 \right|,
 \label{eq:PQR_energy_dependent}
\end{align}
the concurrence can be rewritten as
\begin{align}
 C(s,x)
 =
 \frac{2R(s)(1-x^2)}
 {P(s)(1+x)^2+Q(s)(1-x)^2},
 \qquad
 x=\cos\theta,
 \label{eq:energy_dependent_concurrence}
\end{align}
which has the same form as
Eq.~\eqref{eq:generic_vector_concurrence_PQ},
with the replacements
$P\to P(s)$,
$Q\to Q(s)$,
and
$R\to R(s)$. 
Consequently,
\begin{align}
 \cos\theta_{\rm ext}(s)
 =
 \frac{\sqrt{Q(s)}-\sqrt{P(s)}}
 {\sqrt{P(s)}+\sqrt{Q(s)}}.
 \label{eq:energy_dependent_extremum}
\end{align}
Therefore, the motion of the entanglement extrema is governed by the
energy dependence of the effective chiral combinations $P(s)$ and
$Q(s)$, whereas the weakening of the entanglement is controlled by
the cancellation condition (\ref{eq:interference_zero_condition}).

\begin{figure}[t]
 \centering
 \includegraphics[width=0.55\linewidth]{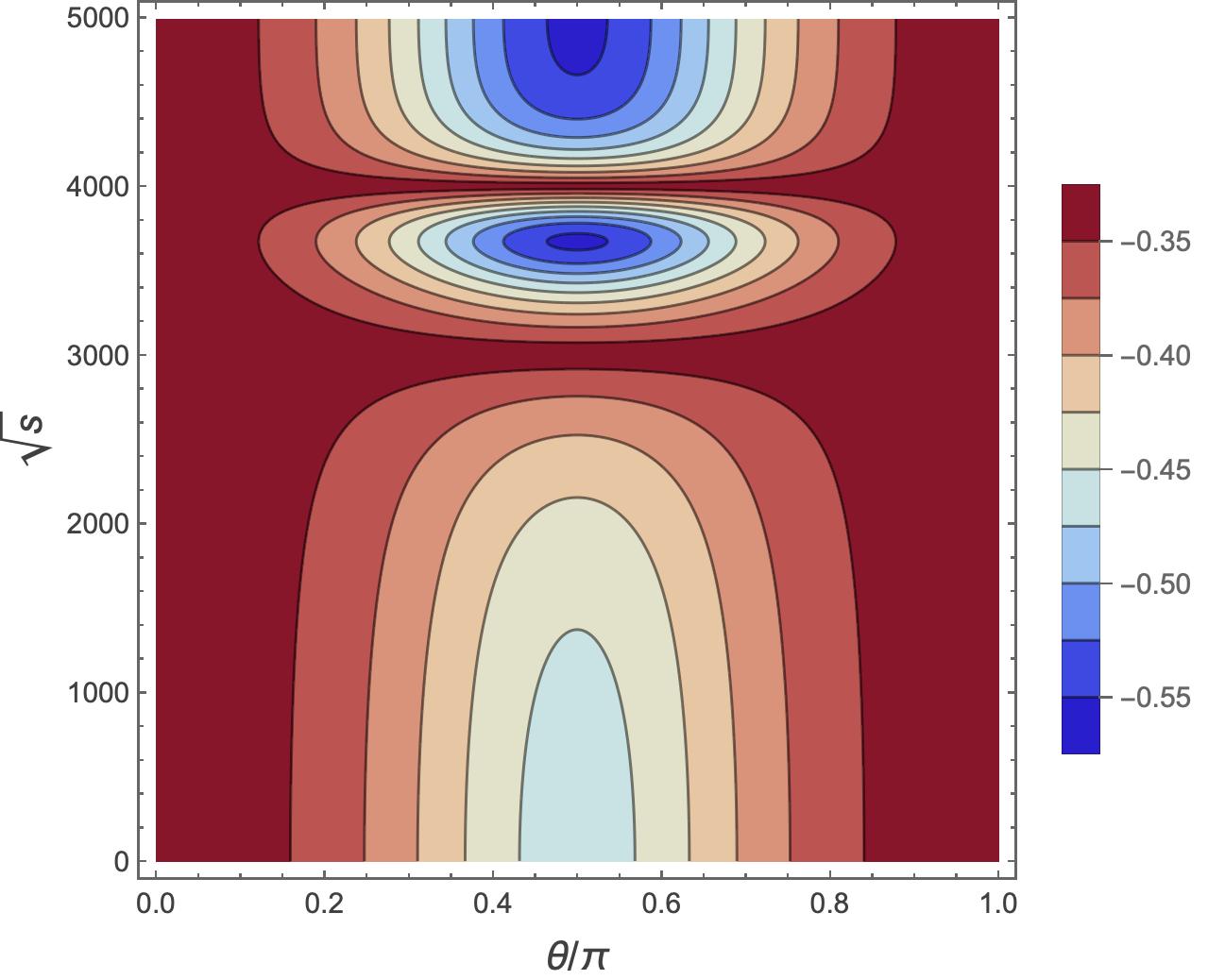}
 \caption{
Contour plot of the entanglement marker obtained from (\ref{eq:interference_entanglement}) 
 with parameters in (\ref{eq:example_chiral_couplings}).
} 
 \label{fig:appendix_chiral_numeric}
\end{figure}

Figure~\ref{fig:appendix_chiral_numeric} shows the contour plot of the
entanglement marker Eq.~(\ref{eq:interference_entanglement}) for the representative 
 parameter set introduced in (\ref{eq:example_chiral_couplings}).  
 In this example, the lepton coupling is vector-like,
$g_L^l=g_R^l$, which implies $P(s)=Q(s)$ and hence
$\theta_{\rm ext}=\pi/2$ at all energies.
The entanglement is strongly suppressed near
 $\sqrt{s}=3~{\rm TeV}$ and $4~{\rm TeV}$, where the off-diagonal
 density-matrix element vanishes as discussed above.

\section{Origin of the disconnected entangled regions in the RS model}
\label{app:RS_angular}

In this appendix, we clarify the origin of the disconnected entangled
regions found in the RS model.
For this purpose, we isolate the contribution from the first KK graviton
and neglect the SM amplitudes.
We also take the massless-top limit, \(m_t=0\), in order to extract the
angular structure associated purely with the spin-2 exchange.

In this limit, the helicity amplitudes can be written schematically as
\begin{equation}
 {\cal M}^{(G)}(s,\theta)
 = A(s)\,{\cal F}(\theta),
\end{equation}
where $A(s)$ is given in (\ref{eq:A}), while \({\cal F}(\theta)\) describes the angular
dependence of the helicity amplitudes.
Since the spin density matrix is normalized by its trace, the common
factor $\rvert A(s)\rvert^2$ cancels.
The normalized density matrix therefore depends only on the scattering
 angle in this simplified limit.
As in Appendix~\ref{app:chiral_vector}, only the
 $\rho_{+-,+-}$, $\rho_{+-,-+}$,
 $\rho_{-+,+-}$, and $\rho_{-+,-+}$ components are nonvanishing.
The normalized density matrix is therefore written as
\begin{equation}
 \rho
 ={1 \over 2}
 \begin{pmatrix}
  1 & 0 & 0 & r(\theta) \\
  0 &  0 & 0 & 0 \\
  0 &  0 & 0 & 0 \\
  r(\theta) & 0 & 0 & 1
 \end{pmatrix},
 \label{eq:rhohat_first_KK}
\end{equation}
where
\begin{equation}
 r(\theta)
 ={2\sin\theta\sin 3\theta \over 2+\cos 2\theta+\cos 4\theta}\,. \label{eq:r_theta}
\end{equation}

For the density matrix in Eq.~\eqref{eq:rhohat_first_KK}, the
entanglement marker reduces to
\begin{equation}
D_{\rm min}(\theta)
 =
 -\frac{1}{3}
 \left(
 1+
2
\left|
r(\theta)
 \right|
 \right).
 \label{eq:marker_first_KK}
\end{equation}
Thus, the entanglement marker returns to its boundary value whenever $r(\theta)=0$.
The denominator in Eq.~\eqref{eq:r_theta} remains positive throughout the physical angular range.
Indeed,
\begin{equation}
 2+\cos 2\theta+\cos 4\theta
 = 8\left(\cos^2\theta-{3 \over 8}\right)^2+{7 \over 8},
\end{equation}
and the polynomial on the right-hand side has no zero for
\(0\leq\theta\leq\pi\).
The numerator vanishes at
\begin{equation}
 \theta
 =
 0,\qquad
 \frac{\pi}{3},\qquad
 \frac{2\pi}{3},\qquad
 \pi,
 \label{eq:angular_nodes}
\end{equation}
as can be easily seen from
\begin{equation}
 \sin\theta\sin 3\theta
 =
 \left(1-\cos^2\theta\right)
 \left(4\cos^2\theta-1\right).
 \label{eq:angular_factorization}
\end{equation}
Thus, the two internal nodes at
\(\theta=\pi/3\) and \(2\pi/3\) divide the physical angular interval into
three separate regions.
Therefore, the separation of the angular regions is entirely determined
by the zeros of the off-diagonal matrix element $r(\theta)$.

The same conclusion follows from the concurrence.
For the density matrix in Eq.~\eqref{eq:rhohat_first_KK}, the
concurrence is given by
\begin{equation}
C(\theta)=\left|r(\theta) \right|
 =
 \left|
 \frac{
 2\sin\theta\sin 3\theta
 }{
 2+\cos 2\theta+\cos 4\theta
 }
 \right|.
 \label{eq:concurrence_first_KK}
\end{equation}
It vanishes at the angles listed in Eq.~\eqref{eq:angular_nodes}, while
it becomes nonzero in the three intervening angular intervals.
Note that the entanglement marker and concurrence satisfy the same
 relation as in (\ref{eq:Dmin_concurrence_relation}).

\begin{figure}[t]
 \centering
 \includegraphics[width=0.55\linewidth]{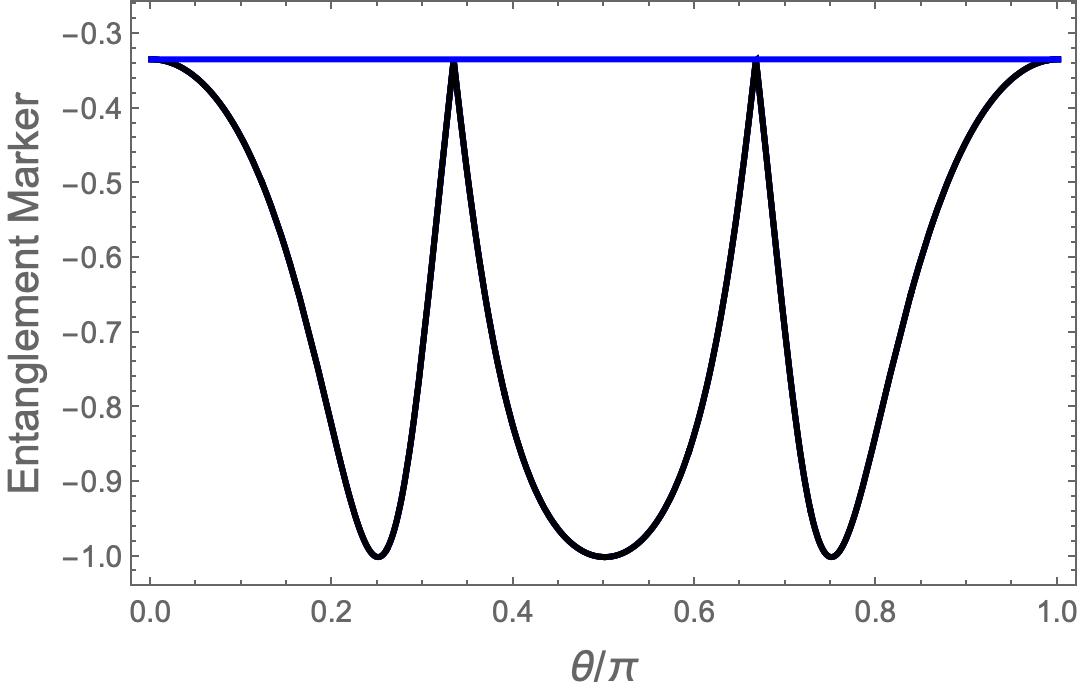}
 \caption{
Entanglement marker obtained by retaining only the first KK graviton
 in the massless-top limit, with the SM amplitudes omitted.
The blue horizontal line denotes the separability bound,
 $D_{\rm min}=-1/3$, while the black curve shows the entanglement marker
 induced by the first KK graviton.
The disconnected entangled regions are already present for a single
 spin-2 KK mode, demonstrating that they originate from the
 characteristic angular dependence of the spin-2 exchange.
} 
 \label{fig:RS_first_KK}
\end{figure}

The numerical result obtained by retaining only the first KK graviton is
shown in Fig.~\ref{fig:RS_first_KK}.
The disconnected angular structure is already present in this
single-mode result.
This demonstrates that the multiple entangled regions are not generated
by interference among different KK gravitons.
Instead, they originate from the higher-order angular dependence
characteristic of the spin-2 exchange, in particular from the
\(\sin 3\theta\) dependence of the helicity coherence.

\vspace{3mm}

\noindent\textbf{Acknowledgments}\\
K.~M. would like to thank Valentin Durupt, Fabio Maltoni, and Olivier Mattelaer for valuable discussions 
 and technical help in numerical computation of spin density matrices.  
The work of M.~A. and K.~M. is supported in part by JSPS Grant-in-Aid for Scientific Research (KAKENHI) 
 under Grant Nos. JP25K07275 and JP24K07032, respectively.
The work of N.~O. is supported in part by the United States Department of Energy Grant
 Nos. DE-SC0012447 and DE-SC0026347.

\end{document}